\documentclass[%
 aip,
 jmp,%
 amsmath,amssymb,
preprint,
]{revtex4-1}

\usepackage{graphicx}
\usepackage{dcolumn}
\usepackage{bm}
\usepackage{verbatim}

\begin{document}

\preprint{AIP/123-QED}

\title[Windrush]{Modification of classical electron transport due to collisions between electrons and fast ions}

\author{B. Appelbe}
\email{bappelbe@ic.ac.uk}
\affiliation{%
 The Centre for Inertial Fusion Studies,  The Blackett Laboratory, Imperial College, London, SW7 2AZ, United Kingdom
}%
\author{M. Sherlock}
\affiliation{%
Lawrence Livermore National Laboratory, P.O. Box 808, Livermore, California 94551-0808, USA
}%

\author{O. El-Amiri}
\affiliation{%
 The Centre for Inertial Fusion Studies,  The Blackett Laboratory, Imperial College, London, SW7 2AZ, United Kingdom
}%

\author{C. Walsh}
\affiliation{%
 The Centre for Inertial Fusion Studies,  The Blackett Laboratory, Imperial College, London, SW7 2AZ, United Kingdom
}%

\author{J. Chittenden}
\affiliation{%
 The Centre for Inertial Fusion Studies,  The Blackett Laboratory, Imperial College, London, SW7 2AZ, United Kingdom
}%

\date{\today}

\begin{abstract}
A Fokker-Planck model for the interaction of fast ions with the thermal electrons in a quasi-neutral plasma is developed. When the fast ion population has a net flux (i.e. the distribution of fast ions is anisotropic in velocity space) the electron distribution function is significantly perturbed from Maxwellian by collisions with the fast ions, even if the fast ion density is orders of magnitude smaller than the electron density. The Fokker-Planck model is used to derive classical electron transport equations (a generalized Ohm's law and a heat flow equation) that include the effects of the electron-fast ion collisions. It is found that these collisions result in a current term in the transport equations which can be significant even when total current is zero. The new transport equations are analyzed in the context of a number of scenarios including $\alpha$ particle heating in ICF and MIF plasmas and ion beam heating of dense plasmas.
\end{abstract}

\maketitle


\section{Introduction}\label{sec:1}
Fast ions in dense, classical plasmas occur in a number of scenarios. Most notable is the $\alpha$ particle heating process which is required to achieve high energy gain in Inertial Confinement Fusion\cite{Lindl_1995} and Magneto-Inertial Fusion\cite{Lindemuth_NF1983,Wurden_MIF2016} schemes. Fast ions are also integral to ion fast ignition schemes\cite{Honrubia_POP2015,Fernandez_NucFus2014}, in which beams of ions are used to ignite compressed DT fuel, and are generated in a range of Z-pinch configurations.\cite{Velikovich_POP2016,Krishnan_IEEE2012}

The effect of the thermal plasma on the fast ions, in the form of ion stopping powers as a function of plasma temperature and density, has been the subject of detailed experimental\cite{Cayzac_Nat2017,Frenje_PRL2015} and theoretical\cite{BPS_2005} studies. However, the effect of the fast ions on the thermal plasma has received less rigorous examination. It is usually assumed that the thermal electrons and ions remain Maxwellian and, therefore, calculation of the ion stopping allows the change in thermal plasma temperature and fluid velocity to be deduced from the conservation of energy and momenturm. This assumption has been verified numerically for the case of collisions between $\alpha$ particles and thermal deuterium and tritium ions in ICF plasmas.\cite{Michta_POP2010,Peigney_POP2014,Sherlock_HEDP2009} However, no such verification appears to exist for the case of fast ion collisions with electrons. Instead, the assumption is usually justified by the argument that $\tau_{ee}$, the timescale for electron equilibration, is much shorter than $\tau_{\alpha e}$, the timescale for stopping of the fast ions due to collisions with electrons.\footnote{We use the subscript $\alpha$ to denote the fast ion population but remind the reader that our model can be generalized to any species of fast ions, not just $\alpha$ particles.} Our work here tests this assumption and demonstrates that it may not always be accurate.

Our approach is to solve the linearized electron kinetic equation using the Fokker-Planck model for particle collisions. The solution to this equation is the perturbation from Maxwellian of the electron distribution. This perturbation is then used to calculate new terms to be included in the generalized Ohm's law, governing the transport of electrical charge, and heat flow equation, governing the transport of thermal energy, of the electrons. Thus, we can evaluate how collisions between electrons and fast ions affect these classical transport equations. Combining Ohm's law with Faraday's law then gives an induction equation that also allows us to determine how these collisions affect the transport of magnetic field in the plasma.

The new terms appearing in the classical electron transport equations are a heat flux and net current arising from the collisions between electrons and fast ions. The net current is the sum of the fast ion current and the electron current induced by electron-fast ion collisions. Interestingly, this term is usually non-zero, meaning that the current induced by electron-fast ion collisions does not exactly cancel the fast ion current, i.e. electron-fast ion collisions result in the generation of current. It is found that this net current increases as the ratio of the charge state of the fast ions to the thermal ions increases in an unmagnetized plasma. For a magnetized plasma, the net current can be dominated by the fast ion current since the Larmor radius of fast ions is much greater than that of the electrons.

The main purposes of this work are
\begin{itemize}
\item[1.] To demonstrate that fast ions, particularly $\alpha$ particles, can perturb electrons from Maxwellian in dense plasmas.
\item[2.] To provide a method for calculating this perturbation using the electron kinetic equation with Fokker-Planck collision operators.
\item[3.] To derive a set of classical electron transport equations that include the effects of fast ions.
\item[4.] To provide estimates of the effects these fast ions can have on the dynamics of both unmagnetized and magnetized plasmas through the transport equations.
\end{itemize}

The contents of this paper are as follows. Section \ref{sec:2} introduces the physical model and outlines some of the major assumptions underpinning our work. Section \ref{sec:3} contains the derivation of the Fokker-Planck model and the classical electron transport equations containing the new terms. Sections \ref{sec:4} and \ref{sec:5} analyze the effects of these terms on the heat flow and magnetic field transport, respectively. Section \ref{sec:6} outlines some effects of the electron-fast ion collisions on the conventional electron transport coefficients and section \ref{sec:7} has some concluding discussions.

\section{Outline of model and some basic assumptions}\label{sec:2}
We consider a plasma with three species, namely, thermal populations of electrons (denoted by subscript $e$) and ions ($i$) and a nonthermal population of fast ions ($\alpha$).\footnote{We use an average ion approximation for a thermal ion population containing both D and T ions.} We assume that the plasma is quasineutral with $n_{e} = Z_{i}n_{i}+Z_{\alpha}n_{\alpha}$. However, we also assume that the fast ion population represents a small fraction of the total plasma ($n_{\alpha}\ll n_{e}$) such that the ion and electron populations can be expected to exhibit fluid-like behaviour. Therefore, we have $n_{e}\approx Z_{i}n_{i}$.

In this work we focus on how the electron population is affected by collisions with the fast ions. For a plasma with similar electron and ion temperatures, $T_{e}\sim T_{i}$, and a much larger fast ion energy, $E_{\alpha}\gg T_{e}$, the fast ions will slow down predominantly due to collisions with electrons since the thermal velocity of the electrons is much greater than the thermal velocity of the ions. Therefore, collisions between fast ions and the thermal ion population are not considered in this work. We assume throughout that the thermal ions (and, correspondingly, the electron-ion fluid) are at rest.

The electron kinetic equation in the rest frame of the ions is given by
\begin{equation}\label{e:2.1}
\frac{\partial f_{e}}{\partial t}+\mathbf{v}\cdot\nabla f_{e}-\frac{e}{m_{e}}\left(\mathbf{E}+\mathbf{v}\times\mathbf{B}\right)\cdot\nabla_{v}f_{e}=\mathcal{C}_{ee}\left(f_{e}\right)+\mathcal{C}_{e i}\left(f_{e},f_{i}\right)+\mathcal{C}_{e\alpha }\left(f_{e},f_{\alpha}\right),
\end{equation}
where $\mathbf{E}$ denotes the electric field in the ion rest frame, $\mathbf{B}$ is the magnetic field, $e$ is the absolute value of the electron charge, $m_e$ is the electron mass and $\mathcal{C}_{ee}$, $\mathcal{C}_{ei}$, $\mathcal{C}_{e\alpha}$ represent Fokker-Planck collision operators for collisions of electrons with electrons, ions and fast ions, respectively.

We begin by expanding the electron distribution function in Cartesian tensors\cite{Johnston_PR1960, Johnston_PR1960_E1, Johnston_PR1960_E2,Shkarofsky_1963}
\begin{equation}\label{e:2.2}
f_{e} = f_{0}+\frac{\mathbf{f_{1}}\cdot\mathbf{v}}{v}+\frac{\underline{\underline{\mathbf{f_{2}}}}:\underline{\underline{\mathbf{vv}}}}{v^{2}}+\ldots,
\end{equation}
in which the terms $f_{0}$, $\mathbf{f_{1}}$, $\underline{\underline{\mathbf{f_{2}}}}$, $\ldots$ depend only on the magnitude of the velocity variable $v$. Throughout this work vectors will be represented in bold (e.g. $\mathbf{v}$) and rank 2 tensors will use a double underline ($\underline{\underline{\boldsymbol{\mu}}}$). By inserting the expansion into \eqref{e:2.1} and taking the first angular moment we obtain the following $\mathbf{f_{1}}$ equation for electrons\cite{Shkarofsky_1966}
\begin{equation}\label{e:2.3}
\frac{\partial \mathbf{f_{1}}}{\partial t}+v\nabla f_{0} -\mathbf{a}\frac{\partial {f_{0}}}{\partial v} -\boldsymbol{\omega}\times\mathbf{f_{1}}+\frac{2}{5}v\nabla\cdot\underline{\underline{\mathbf{f_{2}}}}-\frac{2}{5v^{3}}\frac{\partial}{\partial v}\left(v^{3}\mathbf{a}\cdot\underline{\underline{\mathbf{f_{2}}}}\right)  = \mathbf{C}_{ee1}+\mathbf{C}_{ei1}+\mathbf{C}_{e\alpha1},
\end{equation}
where $\boldsymbol{\omega}=\frac{e}{m_{e}}\mathbf{B}$, $\mathbf{a} = \frac{e}{m_{e}}\mathbf{E}$ and the terms on the RHS represent the first angular moments of the expanded collision operators.

Now, we assume that $f_{0}$ represents a Maxwellian distribution of electrons with density $n_{e}$ and temperature $T_{e}$
\begin{equation}\label{e:2.4}
f_{0} = \frac{n_{e}}{\left(\sqrt{\pi}v_{Te}\right)^{3}}\exp\left(-\frac{v^{2}}{v_{Te}^{2}}\right),
\end{equation}
where $v_{Te} = \sqrt{\frac{2T_{e}}{m_{e}}}$ is the electron thermal velocity. We also assume that $\underline{\underline{\mathbf{f_{2}}}}$ and all higher order terms are negligible. This is equivalent to assuming that electron pressure anisotropies are negligible.

Next, we assume the following ordering of timescales
\begin{equation}\label{e:2.5}
\tau_{\alpha e}\gg\tau_{ei},
\end{equation}
where $\tau_{\alpha e}$ represents the slowing down time of fast ions due to collisions with electrons and $\tau_{ei}$ is the electron-ion collision time. In an ICF hotspot, where the fast ion species are $\alpha$ particles, $\tau_{\alpha e} \sim 10^{-11}\,s$ and $\tau_{ei} \sim 10^{-14}\,s$. This ordering of timescales allows us to assume that the electrons respond instantaneously to collisions with fast ions. Therefore, we can ignore the time dependence in \eqref{e:2.3} and find a steady-state solution. We also assume that
\begin{equation}\label{e:2.6}
v_{\alpha}\tau_{ei}\ll l_{f},
\end{equation}
where $l_{f}$ represents the length-scale of the fluid of electrons and ions. This assumption prevents the fast ion particles from travelling nonlocally in the time that it takes for the electrons to respond to the fast ions. For $\alpha$ particles in an ICF hotspot, we have $v_{\alpha}\tau_{ei} \sim 10^{-7}\,m$ and $l_{f} \sim 10^{-5}\,m$. Finally, in the case where the fast ion species are $\alpha$ particles we need to assume that
\begin{equation}\label{e:2.7}
\tau_{\langle\sigma v\rangle}\gg\tau_{ei},
\end{equation}
where $\tau_{\langle\sigma v\rangle}$ represents the timescale for production of $\alpha$ particles due to DT reactions. For ICF capsules, the duration of the burn phase (the total time over which $\alpha$ particles are produced) is $\sim10^{-10}\,s$. Therefore, it is reasonable to assume that the population of $\alpha$ particles is not changing significantly on timescales of $\tau_{ei} \sim10^{-14}\,s$.

With the assumptions of a Maxwellian $f_{0}$, negligible $\underline{\underline{\mathbf{f_{2}}}}$ and higher order terms, and $\frac{\partial \mathbf{f_{1}}}{\partial t} = 0$, \eqref{e:2.3} becomes
\begin{equation}\label{e:2.8}
v f_{0}\left[\frac{\nabla n_{e}}{n_{e}}+\frac{2}{v_{Te}^{2}}\mathbf{a}+\frac{\nabla T_{e}}{T_{e}}\left(\frac{v^{2}}{v_{Te}^{2}}-\frac{3}{2}\right)\right]=\boldsymbol{\omega}\times\mathbf{f_{1}}+\mathbf{C}_{ee1}+\mathbf{C}_{ei1}+\mathbf{C}_{e\alpha1}.
\end{equation}
All terms on the LHS side of this equation are independent of $\mathbf{f_{1}}$, the perturbation on the electrons. By specifying the macroscopic quantities on the LHS and $\boldsymbol{\omega}$ and the collision terms on the RHS, it is possible to solve this equation for $\mathbf{f_{1}}$ which represents the perturbation of the electrons in response to the imposed conditions. Omitting $\mathbf{C}_{e\alpha1}$ from \eqref{e:2.8} gives an equation that has been used by previous authors to derive classical electron transport coefficients.\cite{Braginskii_1965,Epperlein_1986} Our focus in this work is how the inclusion of the $\mathbf{C}_{e\alpha1}$ term affects the classical electron transport equations.

\section{The electron $\mathbf{f_{1}}$ equation with fast ions}\label{sec:3}
We refer to \eqref{e:2.8} as the electron $\mathbf{f_{1}}$ equation. In order to solve it we need expressions for $\mathbf{C}_{ei1}$, $\mathbf{C}_{ee1}$ and $\mathbf{C}_{e\alpha1}$. For the first two of these terms we make use of well known expressions in the literature.

We assume that the ions have a Maxwellian distribution and are not perturbed by collisions with the fast ions. Since we also assume $T_{i} \sim T_{e}$ (such that $v_{Ti}\ll v_{Te}$) the electron-ion collision term has a simple expression\cite{Shkarofsky_1966}
\begin{equation}\label{e:3.1}
\mathbf{C}_{ei1} = -Y_{ei}n_{i}\frac{1}{v^{3}}\mathbf{f_{1}},
\end{equation}
where $n_{i}$ is the ion density and
\begin{equation}\label{e:3.2}
Y_{ei} = \frac{1}{4\pi}\left(\frac{Z_{i}e^{2}}{\varepsilon_{0}m_{e}}\right)^{2}\ln\Lambda_{ei},
\end{equation}
with $\ln\Lambda_{ei}$ representing the Coulomb logarithm. For convenience, we set $T_{i}=T_{e}$ and $n_{e} = Z_{i}n_{i}$ for the calculations that follow in this work. Electron-electron collisions are more complicated since collisions between perturbed, $\mathbf{f_{1}}$, and unperturbed, $f_{0}$, electrons must be included. The expression for $\mathbf{C}_{ee1}$ is given by \eqref{e:a2.2} in \ref{app:2}. In order to obtain an expression for $\mathbf{C}_{e\alpha 1}$ it is necessary to specify a distribution function for the fast ions. This will be considered next.

We note that similar Fokker-Planck models have been used to study currents driven by ion beams in magnetic confinement fusion devices\cite{Cordey_1979,Fisch_RevModPhys1987} However, these models have not been coupled to the classical transport equations that are usually used to model dense plasmas such as those found in ICF. This is our next step.

\subsection{Including electron-fast ion collisions}\label{sec:3.1}
It is first necessary to specify the form of the fast ion distribution function. For many instances of fast ions in dense plasmas, the fast ions have a significant net flux. For example, thermonuclear reactions create $\alpha$ particles in a DT plasma which are initially produced isotropically with a mean energy of $\approx 3.45\,MeV$ and an energy spectrum which is dominated by thermal broadening.\cite{Appelbe_PPCF2011} Collisions with electrons and ions cause the mean $\alpha$ energy to decrease and the spectrum to broaden. The DT reactivity is very sensitive to the ion temperature.\cite{BoschHale_1992} This causes a significant flux of $\alpha$ particles from hotter to cooler regions of a burning plasma. Given these factors we choose a phenomenological representation of the $\alpha$ particle distribution function that contains an isotropic, $F_{0}$, and anisotropic component, $\mathbf{F_{1}}$, such that
\begin{equation}\label{e:3.1.1}
F_{\alpha} = F_{0}+\frac{\mathbf{F_{1}}\cdot\mathbf{v}}{v}.
\end{equation}
It should be noted that the fast ion distributions are far from a Maxwellian. Therefore, we can have $F_{0}\sim\frac{\mathbf{F_{1}}\cdot\mathbf{v}}{v}$. This is in contrast to the electrons in which the higher order terms in \eqref{e:2.2} represent a perturbation from Maxwellian. However, we do require $F_{0}\sim\left|\mathbf{F_{1}}\right|$ such that $F_{\alpha}$ is non-negative.

Choosing \eqref{e:3.1.1} for the fast ion distribution means that we can use a similar expression for fast ion-electron collisions as that for electron-electron collisions. This expression is given by \eqref{e:a2.2} in \ref{app:2}. We note that we cannot use a simplified collision term like that for electron-ion collisions since the velocity of an $\alpha$ particle can be similar to the thermal velocity of the electrons. The terms in \eqref{e:a2.2} depend on either $f_{0}$ and $\mathbf{F_{1}}$ or $\mathbf{f_{1}}$ and $F_{0}$. Therefore, we can now write \eqref{e:2.8} as
\begin{equation}\label{e:3.1.2}
v f_{0}\left[\frac{\nabla n_{e}}{n_{e}}+\frac{2}{v_{Te}^{2}}\mathbf{a}+\frac{\nabla T_{e}}{T_{e}}\left(\frac{v^{2}}{v_{Te}^{2}}-\frac{3}{2}\right)\right]+\mathbf{C}_{e\alpha1}^{01}=\boldsymbol{\omega}\times\mathbf{f_{1}}+\mathbf{C}_{ee1}+\mathbf{C}_{ei1}+\mathbf{C}_{e\alpha1}^{10},
\end{equation}
where all terms on the LHS are independent of $\mathbf{f_{1}}$. The superscripts in the $\mathbf{C}_{e\alpha1}^{ij}$ collision terms denote that the term is dependent on $f_{i}$ and $F_{j}$. Further algebraic manipulation of \eqref{e:3.1.2} results in the following linear integro-differential equation
\begin{eqnarray}\label{e:3.1.3}
&&g_{1}\left(v\right)\frac{\partial^{2} \mathbf{f_{1}}}{\partial v^{2}}+g_{2}\left(v\right)\frac{\partial \mathbf{f_{1}}}{\partial v}+g_{3}\left(v\right)\mathbf{f_{1}}+g_{4}\left(v\right)\int_{0}^{v}v^{5}\mathbf{f_{1}}dv\nonumber\\
&&+g_{5}\left(v\right)\int_{0}^{v}v^{3}\mathbf{f_{1}}dv+g_{6}\left(v\right)\int_{v}^{\infty}\mathbf{f_{1}}dv+\omega\tau_{ei}\mathbf{b}\times\mathbf{f_{1}}=\mathbf{g_{7}}\left(v\right),
\end{eqnarray}
where $\omega\tau_{ei}$ represents the Hall parameter and $\tau_{ei}$ is the electron-ion collision time
\begin{equation}\label{e:3.1.4}
\tau_{ei} = \frac{3\sqrt{\pi}v_{Te}^{3}}{4Y_{ei}n_{i}}.
\end{equation}
The terms $g_{1-6}$  are scalar functions of the isotropic components of the electron and $\alpha$ distributions ($f_{0}$ and $F_{0}$). They do not have a directional component. Expressions for these terms are listed in \eqref{e:a2.7}-\eqref{e:a2.12} of \ref{app:2}.

However, the $\mathbf{g_{7}}$ term does have a directional component
\begin{eqnarray}\label{e:3.1.5}
\mathbf{g_{7}}\left(v\right)&=& \tau_{ei} v f_{0}\left[\frac{\nabla n_{e}}{n_{e}}+\frac{2}{v_{Te}^{2}}\mathbf{a}+\frac{\nabla T_{e}}{T_{e}}\left(\frac{v^{2}}{v_{Te}^{2}}-\frac{3}{2}\right)\right]
-\tau_{ei} Y_{e\alpha}f_{0}\left[4\pi\frac{m_{e}}{m_{\alpha}}\mathbf{F_{1}}
+\frac{4}{5v_{Te}^{4}}vI_{3}^{\alpha 1}\right.\nonumber\\
&&\left.+\frac{2}{5v_{Te}^{2}}\left(2\frac{v^{2}}{v_{Te}^{2}}-\frac{10}{3}+\frac{5m_{e}}{3m_{\alpha}}\right)\frac{1}{v}J_{-2}^{\alpha 1}-\frac{2}{15v_{Te}^{2}}\left(-5+10\frac{m_{e}}{m_{\alpha}}\right)\frac{1}{v}I_{1}^{\alpha 1}\right].\nonumber
\end{eqnarray}
The $I^{\alpha}$ and $J^{\alpha}$ functions used here are defined by \eqref{e:a2.5} and \eqref{e:a2.6} in \ref{app:2}.

As can be seen, the $\mathbf{g_{7}}$ term is a linear function of $\nabla n_{e}$, $\nabla T_{e}$, $\mathbf{E}$ and $\mathbf{F_{1}}$. These terms are the thermodynamic driving forces acting on the electrons. Transport processes occur when these terms have non-zero values. For a given thermodynamic driving force, \eqref{e:3.1.3} can be solved for $\mathbf{f_{1}}$ which determines the electron response to the force. The solution to \eqref{e:3.1.3} has been widely studied for non-zero values of $\nabla n_{e}$, $\nabla T_{e}$, $\mathbf{E}$.\cite{Braginskii_1965,Epperlein_1986} Our goal here is to study the solution to \eqref{e:3.1.3} for non-zero values of $\mathbf{F_{1}}$. This is achieved by specifying values for $F_{0}$ and $\mathbf{F_{1}}$ (with all other thermodynamic forces set to zero) and solving \eqref{e:3.1.3} (using the procedure outlined in \eqref{app:3}) for $\mathbf{f_{1}}$. The following quantities can then be calculated
\begin{eqnarray}
\boldsymbol{\xi}_{e\alpha} &=& -\frac{4\pi e}{3}\int_{0}^{\infty}v^{3}\mathbf{f_{1}}dv,\label{e:3.1.6}\\
\boldsymbol{\zeta}_{e\alpha} &=& \frac{2\pi m_{e}}{3}\int_{0}^{\infty}v^{5}\mathbf{f_{1}}dv.\label{e:3.1.7}
\end{eqnarray}
The expression in \eqref{e:3.1.6} is the current of electrons induced by collisions with the $\alpha$ particles while \eqref{e:3.1.7} represents the corresponding electron heat flow.

\subsection{Obtaining the transport equations}\label{sec:3.2}
The linearity of \eqref{e:3.1.3} with respect to the thermodynamic forces allows us to write down the following transport relations\cite{Shkarofsky_POF1963}
\begin{eqnarray}
\mathbf{j}_{e}^{0}&=& \underline{\underline{\boldsymbol{\sigma}}}\cdot\left(\mathbf{E}+\frac{T_{e}}{e n_{e}}\nabla n_{e}\right)+\underline{\underline{\boldsymbol{\tau}}}\cdot\nabla T_{e}+\boldsymbol{\xi}_{e\alpha},\label{e:3.2.1}\\
\mathbf{q}_{e} &=& -\underline{\underline{\boldsymbol{\mu}}}\cdot\left(\mathbf{E}+\frac{T_{e}}{e n_{e}}\nabla n_{e}\right)-\underline{\underline{\boldsymbol{K}}}\cdot\nabla T_{e}+\boldsymbol{\zeta}_{e\alpha},\label{e:3.2.2}
\end{eqnarray}
where $\underline{\underline{\boldsymbol{\sigma}}}$ is the electric conductivity tensor, $\underline{\underline{\boldsymbol{\tau}}}$ is the thermoelectric tensor, $\underline{\underline{\boldsymbol{\mu}}}$ is the energy conductivity tensor and $\underline{\underline{\boldsymbol{K}}}$ is the thermal diffusion tensor. These can be calculated from the electron $\mathbf{f_{1}}$ equation.\cite{Epperlein_1986} The term $\mathbf{j}_{e}^{0}$ represents the current of electrons flowing in the ion rest frame as a result of the thermodynamic forces while $\mathbf{q}_{e}$ is the corresponding electron heat flow.

We can calculate the current of $\alpha$ particles from \eqref{e:3.1.1}
\begin{equation}\label{e:3.2.3}
\mathbf{j}_{\alpha}^{0} = n_{\alpha}Z_{\alpha}e\langle\mathbf{v}_{\alpha}\rangle = \frac{4\pi Z_{\alpha}e}{3}\int_{0}^{\infty}v^{3}\mathbf{F_{1}}dv,
\end{equation}
where $\langle\mathbf{v}_{\alpha}\rangle$ is the drift velocity of $\alpha$ particles. Now, adding $\mathbf{j}_{\alpha}^{0}$ to both sides of \eqref{e:3.2.1} and multiplying by the inverse of the electric conductivity tensor gives
\begin{equation}\label{e:3.2.4}
\mathbf{E}+\frac{T_{e}}{e n_{e}}\nabla n_{e} = \underline{\underline{\boldsymbol{\sigma}}}^{-1}\cdot\left(\mathbf{j}_{T}-\mathbf{j}_{e\alpha}^{0}\right)-\underline{\underline{\boldsymbol{\sigma}}}^{-1}\underline{\underline{\boldsymbol{\tau}}}\cdot\nabla T_{e},
\end{equation}
Here we have introduced the total current $\mathbf{j}_{T}=\mathbf{j}_{e}^{0}+\mathbf{j}_{\alpha}^{0}$ (this will be the same in the lab frame and the ion rest frame) and $\mathbf{j}_{e\alpha}^{0} = \boldsymbol{\xi}_{e\alpha}+\mathbf{j}_{\alpha}^{0}$ which represents the total current that would flow if electron-$\alpha$ collisions were the only transport process occurring.

The Onsager reciprocal relations give the following relationship between transport coefficients\cite{HochstimMassel_1969}
\begin{equation}\label{e:3.2.5}
\underline{\underline{\boldsymbol{\tau}}}=\frac{1}{T_{e}}\underline{\underline{\boldsymbol{\mu}}}-\frac{3}{2e}\underline{\underline{\boldsymbol{\sigma}}}.
\end{equation}
Applying this to \eqref{e:3.2.4} and using $P_{e}=n_{e}T_{e}$ gives
\begin{equation}\label{e:3.2.6}
\mathbf{E}=-\frac{\nabla P_{e}}{e n_{e}}  +\underline{\underline{\boldsymbol{\sigma}}}^{-1}\cdot\left(\mathbf{j}_{T}-\mathbf{j}_{e\alpha}^{0}\right)-\frac{1}{e}\underline{\underline{\boldsymbol{\beta}}} \cdot\nabla T_{e},
\end{equation}
while combining \eqref{e:3.2.4} with \eqref{e:3.2.2} gives
\begin{equation}\label{e:3.2.7}
\mathbf{q}_{e}= -\underline{\underline{\boldsymbol{\kappa}}}\cdot\nabla T_{e}-\frac{T_{e}}{e}\left(\underline{\underline{\boldsymbol{\beta}}}+\frac{5}{2}\underline{\underline{\mathbf{I}}}\right)\cdot\left(\mathbf{j}_{T}-\mathbf{j}_{e\alpha}^{0}\right)+\boldsymbol{\zeta}_{e\alpha},
\end{equation}
where $\underline{\underline{\boldsymbol{\kappa}}}$ and $\underline{\underline{\boldsymbol{\beta}}}$ represent thermal conductivity and thermoelectric tensors.

Equations \eqref{e:3.2.6} and \eqref{e:3.2.7} are the classical electron transport relations including electron-$\alpha$ collisions. The first equation is Ohm's law, containing terms which affect the electron momentum. It is usually combined with Faraday's law to obtain the induction equation which governs the evolution of the $\mathbf{B}$ field. The second equation is the heat flow equation, governing the rate at which electron thermal energy is transported. We have used \eqref{e:3.2.4} to eliminate explicit dependence of the heat flow on the electric field. However, the $\underline{\underline{\boldsymbol{\kappa}}}$ and $\underline{\underline{\boldsymbol{\beta}}}$ transport coefficients are dependent on the conductivity $\underline{\underline{\boldsymbol{\sigma}}}$ as follows
\begin{eqnarray}
\underline{\underline{\boldsymbol{\kappa}}} &=& -\frac{1}{T_{e}}\underline{\underline{\boldsymbol{\mu}}}\underline{\underline{\boldsymbol{\sigma}}}^{-1} \underline{\underline{\boldsymbol{\mu}}}+\frac{3}{2e}\underline{\underline{\boldsymbol{\mu}}}+\underline{\underline{\boldsymbol{K}}},\label{e:3.2.8}\\
\underline{\underline{\boldsymbol{\beta}}} &=& \frac{e}{T_{e}}\underline{\underline{\boldsymbol{\mu}}}\underline{\underline{\boldsymbol{\sigma}}}^{-1}-\frac{5}{2}\underline{\underline{\mathbf{I}}}.\label{e:3.2.9}
\end{eqnarray}
Therefore, these coefficients account for the return current of electrons which is necessary to maintain the condition $\nabla\cdot\mathbf{j}_{T} = 0$ in quasi-neutral plasmas.\cite{Spitzer_1953} The $\mathbf{j}_{e\alpha}^{0}$ term in \eqref{e:3.2.7} represents heat flow due to the return current of electrons generated by electron-fast ion collisions. As shown in section \ref{sec:4.1} and fig. \ref{f:4} this heat flow is in the opposite direction to $\boldsymbol{\zeta}_{e\alpha}$ for collisions between electrons and $\alpha$ particles in a DT plasma.

The transport coefficients are tensors whose components are defined with respect to the magnetic field unit vector $\mathbf{b}$. Using $\mathbf{s}$ to denote the driving force  (e.g. $\nabla T_{e}$), the components of a tensor $\underline{\underline{\boldsymbol{\varphi}}}$ (e.g. $\underline{\underline{\boldsymbol{\kappa}}}$) are defined as follows
\begin{equation}\label{e:3.2.10}
\underline{\underline{\boldsymbol{\varphi}}}\cdot\mathbf{s} = \varphi_{\|}\left(\mathbf{b}\cdot\mathbf{s}\right)\mathbf{b}+\varphi_{\bot}\mathbf{b}\times\left(\mathbf{s}\times\mathbf{b}\right)+\varphi_{\wedge}\mathbf{b}\times\mathbf{s}.
\end{equation}
In the case of conductivity $\underline{\underline{\boldsymbol{\sigma}}}$ we need to also consider the components of the inverse tensor
\begin{equation}\label{e:3.2.11}
\underline{\underline{\boldsymbol{\varphi}}}^{-1}\cdot\mathbf{s} = \frac{1}{\varphi_{\|}}\left(\mathbf{b}\cdot\mathbf{s}\right)\mathbf{b}+\frac{\varphi_{\bot}}{\varphi_{\bot}^{2}+\varphi_{\wedge}^{2}}\mathbf{b}\times\left(\mathbf{s}\times\mathbf{b}\right)-\frac{\varphi_{\wedge}}{\varphi_{\bot}^{2}+\varphi_{\wedge}^{2}}\mathbf{b}\times\mathbf{s}.
\end{equation}

\subsection{A representation for the $\alpha$ distribution}\label{sec:3.3}
We have shown in \ref{sec:3.1} and \ref{sec:3.2} that by assuming the distribution of fast ions can be represented by an isotropic and anisotropic component, \eqref{e:3.1.1}, we can obtain a set of classical electron transport equations which include the effects of fast ion-electron collisions, \eqref{e:3.2.6}-\eqref{e:3.2.7}. Further analysis of these transport equations requires that we specify the functions $F_{0}$ and $\mathbf{F_{1}}$. Since our goal in this work is to provide estimates of the effects of fast ions on electron transport we can choose simple expressions for these functions. In particular, we can assume that the fast ions are mono-energetic and use the following expressions
\begin{eqnarray}
F_{0} &=& n_{\alpha}\frac{\delta\left(v-v_{\alpha}\right)}{4\pi v^{2}},\label{e:3.3.1}\\
\mathbf{F_{1}} &=& \gamma_{\alpha}n_{\alpha}\frac{\delta\left(v-v_{\alpha}\right)}{4\pi v^{2}}\mathbf{\hat{u}_{\alpha}},\label{e:3.3.2}
\end{eqnarray}
where $v_{\alpha}=\sqrt{2E_{\alpha}/m_{\alpha}}$ is the velocity of the mono-energetic fast ions. We have here introduced three parameters to describe the mono-energetic distribution. These are the density of fast ions, $n_{\alpha}$, the energy of the particles, $E_{\alpha}$, and $\gamma_{\alpha}\in\left[0,1\right]$, which is a measure of the anisotropy of the $\alpha$ distribution. The drift velocity of the population of $\alpha$ particles is now given by
\begin{equation}\label{e:3.3.3}
\langle\mathbf{v}_{\alpha}\rangle = \frac{1}{3}\gamma_{\alpha}v_{\alpha}\mathbf{\hat{u}_{\alpha}}.
\end{equation}
Calculation of the $I^{\alpha}$ and $J^{\alpha}$ functions, given by \eqref{e:a2.5} and \eqref{e:a2.6}, become particularly straightforward due to the Dirac delta functions in \eqref{e:3.3.1} and \eqref{e:3.3.2}.

\begin{figure}
\begin{center}%
\includegraphics*[width=0.95\columnwidth]{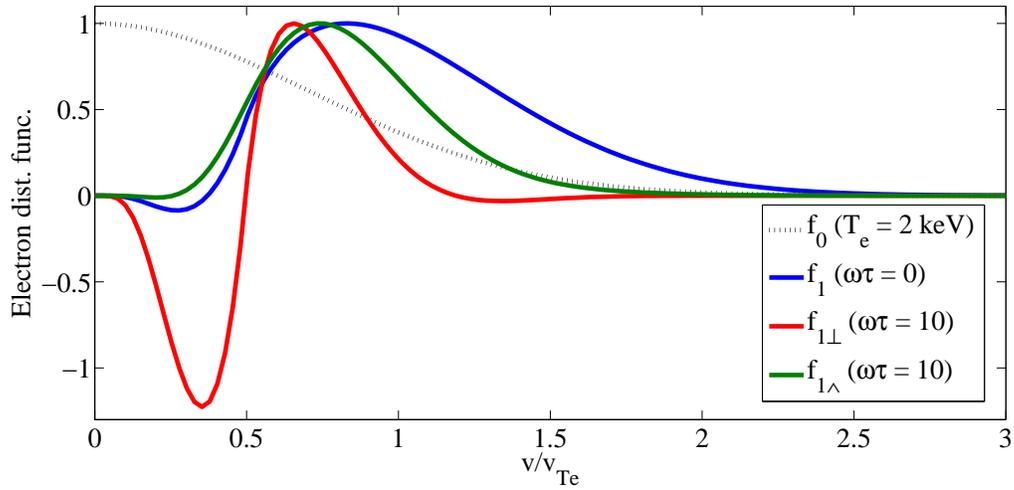}%
\vspace{-1em}
\end{center}
\caption[]{The electron $\mathbf{f_{1}}$ solution arising from electron-$\alpha$ collisions for both an unmagnetized (blue) and magnetized plasma (red and green) with $T_{e} = 2\,keV$ and $E_{\alpha} = 3.45\,MeV$. The Maxwellian component of the electron distribution function is also shown for comparison.} \label{f:1}
\end{figure}

With $F_{\alpha}$ now defined, we can calculate the solution to \eqref{e:3.1.5} and the resulting $\boldsymbol{\xi}_{e\alpha}$ and $\boldsymbol{\zeta}_{e\alpha}$ values. As an example, figure \ref{f:1} illustrates the $\mathbf{f_{1}}$ component of the electron distribution function for an $\alpha$ particle with energy of $3.45\,MeV$ in a DT plasma with $T_{e} = 2\,keV$. This means that the $\alpha$ particle velocity $v_{\alpha} = 0.48v_{Te}$. As is clear in figure \ref{f:1}, the electron-$\alpha$ collisions are perturbing electrons with a velocity several times larger than $v_{\alpha}$ in both unmagnetized and magnetized plasmas.  The shape of the $\mathbf{f_{1}}$ function is independent of $n_{e}$, $n_{\alpha}$ and $\langle\mathbf{v}_{\alpha}\rangle$.

Since both $\mathbf{f_{1}}$ and $\mathbf{F_{1}}$ are proportional to the fast ion particle flux, $n_{\alpha}\langle v_{\alpha}\rangle$ (as defined by the parameter $\gamma_{\alpha}$), it follows that the terms $\mathbf{j}_{\alpha}^{0}$, $\boldsymbol{\xi}_{e\alpha}$ and $\boldsymbol{\zeta}_{e\alpha}$ are also all proportional to the fast ion flux. In addition, $\mathbf{j}_{e\alpha}^{0}$ and $\boldsymbol{\zeta}_{e\alpha}$ will be independent of $n_{e}$, assuming that $n_{\alpha}\ll n_{e}$. Therefore, for a mono-energetic population of fast ions, the current heat flow generated by electron-fast ion collisions will be a function of the ratio of velocities, $\frac{v_{\alpha}}{v_{Te}}$, electron magnetization, $\omega\tau$, and directly proportional to fast ion flux.

\begin{figure}
\begin{center}%
\includegraphics*[width=0.95\columnwidth]{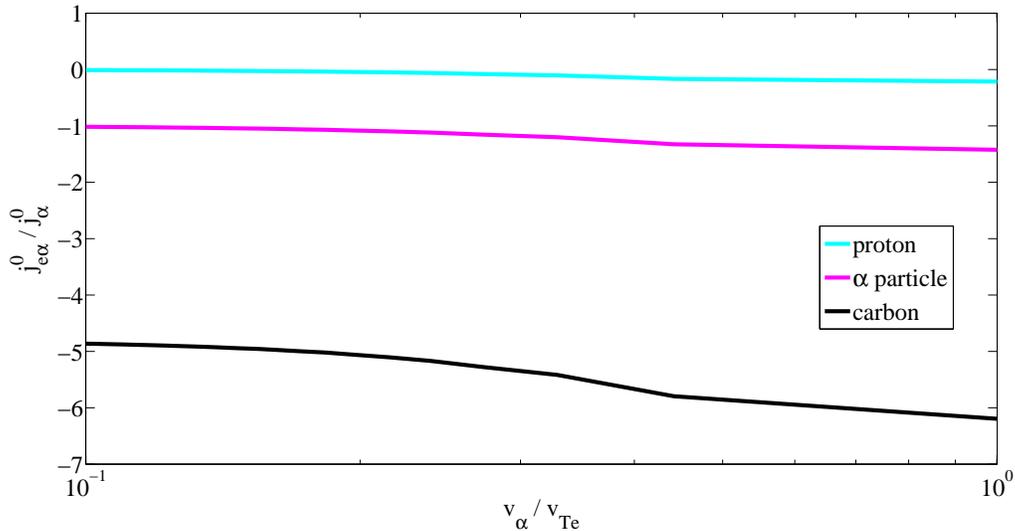}%
\vspace{-1em}
\end{center}
\caption[]{The ratio of the net current induced by fast ion-electron collisions, $j^{0}_{e\alpha}$, to the current of fast ions, $j_{\alpha}^{0}$, as a function of the ratio of fast ion particle velocity to electron thermal velocity for fast protons, $\alpha$ particles and carbon ions (all fully ionized) in an unmagnetized DT plasma.} \label{f:2}
\end{figure}

It has been shown previously\cite{Ohkawa_1970,Fisch_RevModPhys1987} that, for unmagnetized plasmas in the limit of $\frac{v_{\alpha}}{v_{Te}}\rightarrow 0$, the net current induced by electron-fast ion collisions is related to the charge state of the fast ion and thermal ion populations by
\begin{equation}\label{e:3.3.4}
\frac{\mathbf{j}_{e\alpha}^{0}}{\mathbf{j}_{\alpha}^{0}} \approx 1-\frac{Z_{\alpha}}{Z_{i}}.
\end{equation}
The results shown in fig. \ref{f:2} for an unmagnetized plasma are well approximated by this relationship. These results demonstrate that collisions between fast ions and electrons will generate a net current in plasma, even for an unmagnetized plasma in the quasi-neutral approximation.

\begin{figure}
\begin{center}%
\includegraphics*[width=0.95\columnwidth]{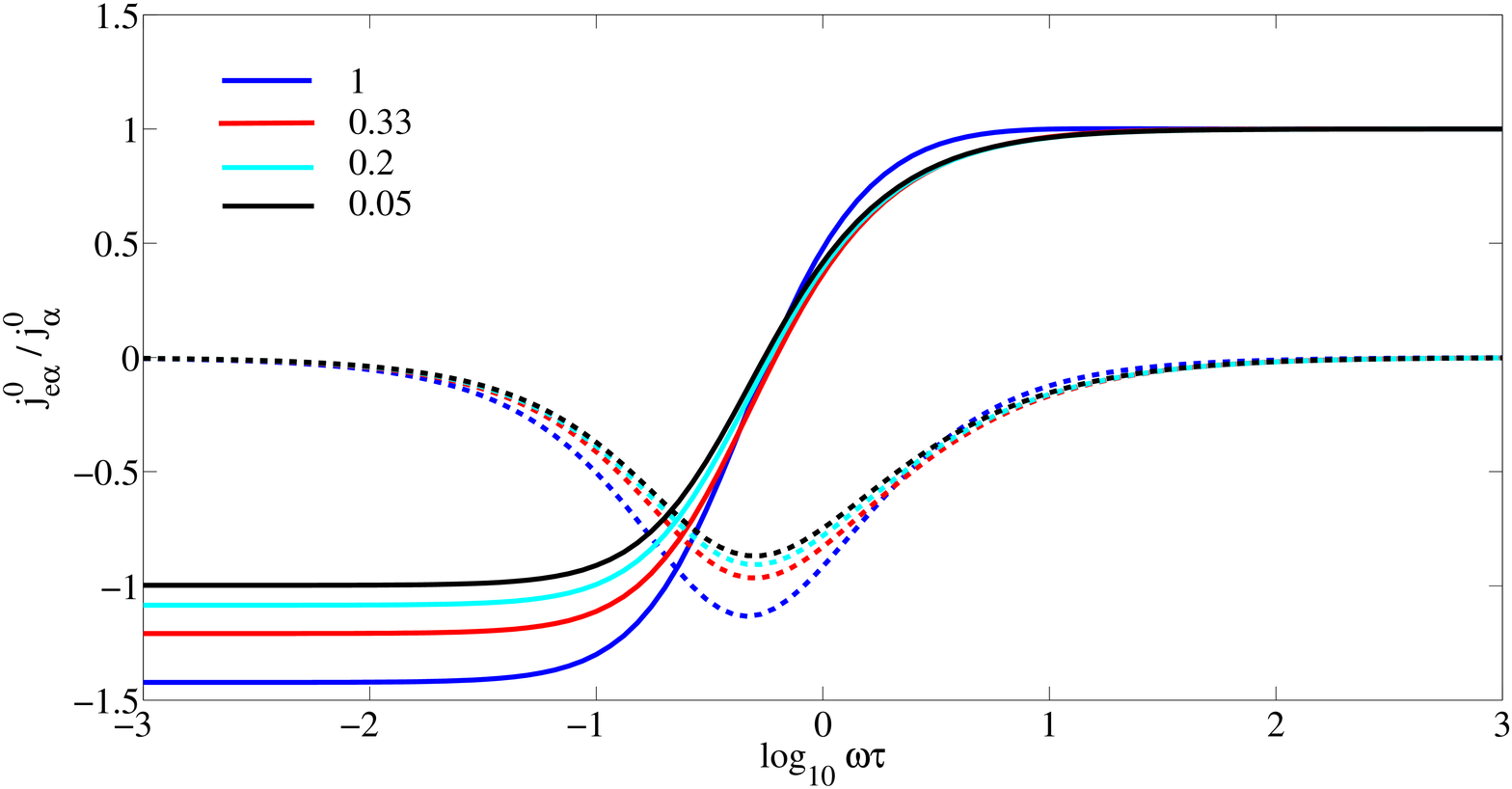}%
\hfil
\includegraphics*[width=0.95\columnwidth]{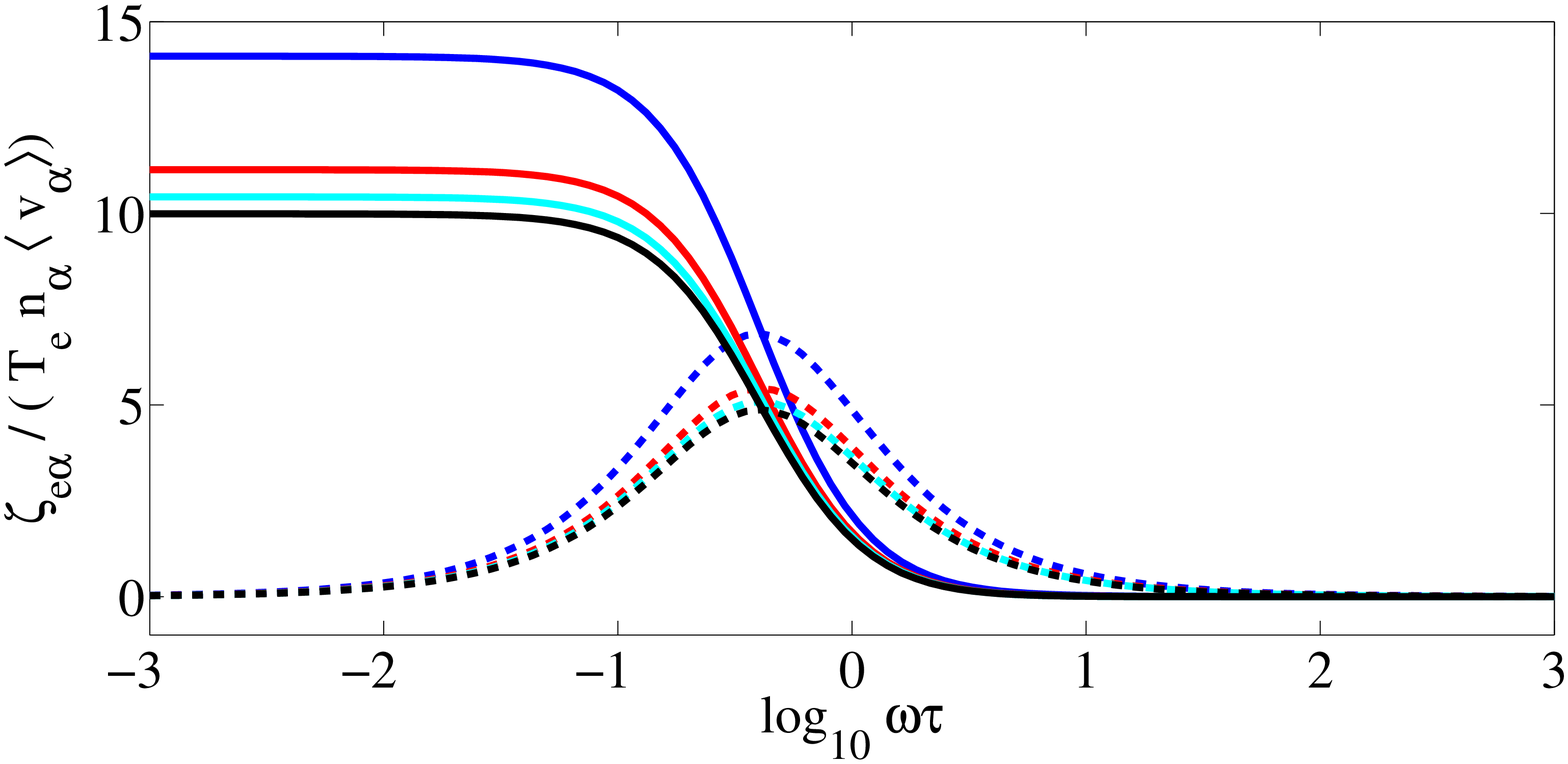}%
\vspace{-1em}
\end{center}
\caption[]{Top: The ratio of net current due to $\alpha$-electron collisions to $\alpha$ current as a function of $\omega\tau$ for several values of $v_{\alpha}/v_{Te}$. The full lines denote components in the $\bot$ direction (i.e. parallel to $\alpha$ flux) and the dashed lines denote components in the $\wedge$ direction (orthogonal to $\alpha$ flux and $\mathbf{B}$ field). Bottom: A similar plot for the heat flows induced by $\alpha$-electron collisions. This quantity is normalized by the $\alpha$ particle flux and electron temperature.} \label{f:3}
\end{figure}

Magnetized plasmas also display interesting behaviour. This is illustrated in fig. \ref{f:3} for the case of $\alpha$ particles in a DT plasma. Most notably, we see that as we move from unmagnetized plasma, $\omega\tau\rightarrow 0$, to highly magnetized, $\omega\tau\rightarrow \infty$, the net current induced by $\alpha$-electron collisions, $\left(\mathbf{j}^{0}_{e\alpha}\right)_{\bot}$, reverses direction. This is because the electron current is suppressed at high magnetization ($\boldsymbol{\xi}_{e\alpha}\rightarrow 0$ as $\omega\tau\rightarrow \infty$) and so $\left(\mathbf{j}^{0}_{e\alpha}\right)_{\bot} \approx \mathbf{j}^{0}_{\alpha}$.  Meanwhile, for plasmas with intermediate magnetization, $\omega\tau \sim 1$, there is a significant induced current orthogonal to the $\alpha$ flux, $\left(\mathbf{j}^{0}_{e\alpha}\right)_{\wedge}$. This current is directed such that, due to Lenz's law, it opposes the magnetic field inducing it.

The bottom diagram in fig. \ref{f:3} shows the effects of magnetization on the $\boldsymbol{\zeta}_{e\alpha}$ term. Here we have normalized $\boldsymbol{\zeta}_{e\alpha}$ by the electron temperature and the $\alpha$ particle flux. The component parallel to the $\alpha$ particle flux, $\left(\boldsymbol{\zeta}_{e\alpha}\right)_{\bot}$, monotonically drops to $0$ as $\omega\tau$ increases. This is because, effectively, the increasing magnetization reduces the ability of electrons to transport thermal energy. The $\left(\boldsymbol{\zeta}_{e\alpha}\right)_{\wedge}$ component becomes significant at intermediate magnetization values. A polynomial fit to the data shown in fig. \ref{f:3} is given in \ref{app:4}.

\section{Heat flow effects of electron-fast ion collisions}\label{sec:4}
The electron heat flow equation is given by \eqref{e:3.2.7}. The fast ions are responsible for two terms in this equation. First, the $\mathbf{j}_{e\alpha}^{0}$ term, which is a thermoelectric effect in which the current generated by electron-fast ion collisions transports thermal energy and, second, the $\boldsymbol{\zeta}_{e\alpha}$ term which is the heat flow generated directly by the collisions. We now examine these terms in an unmagnetized and magnetized plasma.

\subsection{Heat flow in an unmagnetized plasma}\label{sec:4.1}
\begin{figure}
\begin{center}%
\includegraphics*[width=0.95\columnwidth]{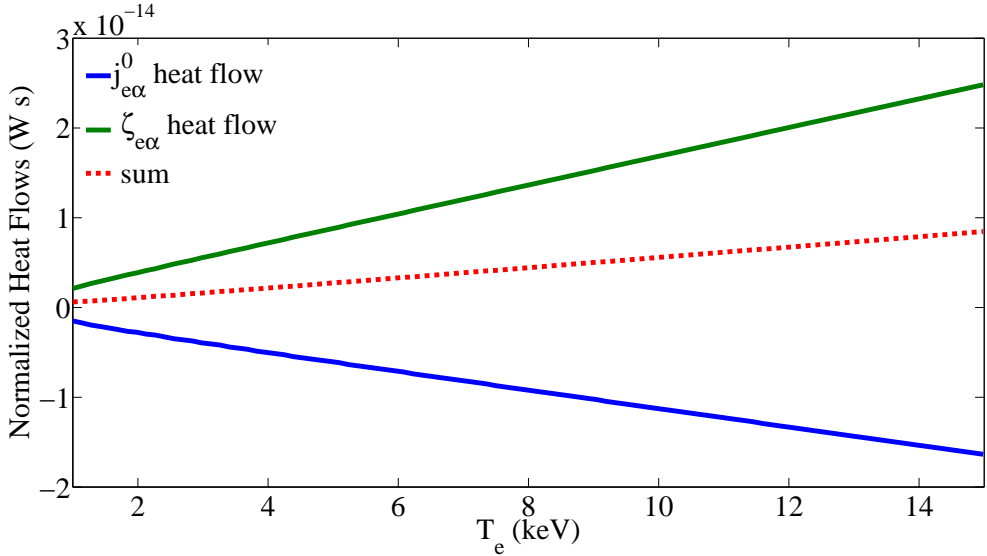}%
\vspace{-1em}
\end{center}
\caption[]{Normalized heat flow due to electron-fast ion collisions where the fast ions are a mono-energetic population of $\alpha$ particles with $E_{\alpha} = 3.45\,MeV$. The heat flow values have been normalized by the $\alpha$ flux $n_{\alpha}\langle v_{\alpha}\rangle$.} \label{f:4}
\end{figure}

For an unmagnetized plasma we have $\mathbf{B} = 0$ and $\mathbf{j}_{T}=0$ and so the heat flow resulting from fast ion-electron collisions can be expressed as
\begin{equation}\label{e:4.1.1}
\mathbf{q}_{e}= \frac{T_{e}}{e}\left(\beta_{\|}+\frac{5}{2}\right)\mathbf{j}_{e\alpha}^{0}+\boldsymbol{\zeta}_{e\alpha},
\end{equation}
where $\beta_{\|}$ is the unmagnetized thermoelectric coefficient. Note that, although the total current in the plasma is zero, the current arising from fast ion-electron collisions, $\mathbf{j}_{e\alpha}^{0}$, can contribute to heat flow. To investigate \eqref{e:4.1.1} we choose a mono-energetic population of $\alpha$ particles with $E_{\alpha} = 3.45\,MeV$ in a DT plasma. In fig. \ref{f:4} we plot the heat flow terms normalized by the $\alpha$ particle flux, as a function of $T_{e}$. As can be seen, the heat flows due to $\mathbf{j}_{e\alpha}^{0}$ and $\boldsymbol{\zeta}_{e\alpha}$ point in opposite directions, with $\boldsymbol{\zeta}_{e\alpha}$ parallel to the direction of $\alpha$ flux and $\mathbf{j}_{e\alpha}^{0}$ anti-parallel. The $\boldsymbol{\zeta}_{e\alpha}$ contribution is larger and so the heat flow due to electron-$\alpha$ collisions is parallel to $\alpha$ flux .

Recent indirect-drive ICF experiments on the NIF have produced yields of $1.9\times 10^{16}$ DT neutrons (and $\alpha$ particles).\cite{LePape_PRL2018} Experimental measurements indicate that these reactions occurred in a hotspot of radius $\sim 30\,\mu m$ and over a time period of $\sim 150\,ps$. Given that the slowing down time of the $\alpha$ particles is $\sim 10\,ps$ we can estimate that there is an average density of $\sim 10^{28}\,m^{-3}$ $\alpha$ particles with energy in the $MeV$ range during the burn pulse. Computer simulations suggest that $\alpha$ populations in the hotspot can have drift velocities $\langle v_{\alpha}\rangle \sim10^{6}\,m\,s^{-1}$ giving $\alpha$ fluxes $n_{\alpha}\langle v_{\alpha}\rangle\sim 10^{34}\,m^{-2}\,s^{-1}$. Therefore, using fig. \ref{f:4} that heat flows in the range of $10^{19}-10^{20}\,W\,m^{-2}$ could be occurring in the hotspot due to electron-$\alpha$ collisions. Equilibrium hotspot models\cite{Lindl_1995} suggest that heat flow losses due to thermal conduction ($\kappa_{\|}\nabla T_{e}$) from the hotspot to the cold fuel are of a similar order of magnitude. Therefore, we conclude that the heat flows due to electron-$\alpha$ collisions could make a significant contribution to the transport of thermal energy in these ICF experiments. Further investigation of this will require incorporating the heat flow equation \eqref{e:4.1.1} into integrated simulation codes in which the $\alpha$ fluxes and particle energies in different regions of the hotspot can be accurately calculated.

\subsection{Heat flow in a magnetized plasma}\label{sec:4.2}
For simplicity we assume a magnetized plasma with a uniform $\mathbf{B}$ field and, therefore, $\mathbf{j}_{T} = 0$. Setting the fast ion flux orthogonal to the $\mathbf{B}$ field means that the components of the induced heat flow can be expressed as
\begin{eqnarray}
q_{e\bot}&=& \frac{T_{e}}{e}\left(\beta_{\bot}+\frac{5}{2}\right)\left(j^{0}_{e\alpha}\right)_{\bot}-\frac{T_{e}}{e}\beta_{\wedge}\left(j^{0}_{e\alpha}\right)_{\wedge}+\left(\zeta_{e\alpha}\right)_{\bot},\label{e:4.2.2}\\
q_{e\wedge}&=& \frac{T_{e}}{e}\beta_{\wedge}\left(j^{0}_{e\alpha}\right)_{\bot}+\frac{T_{e}}{e}\left(\beta_{\bot}+\frac{5}{2}\right)\left(j^{0}_{e\alpha}\right)_{\wedge}+\left(\zeta_{e\alpha}\right)_{\wedge},\label{e:4.2.3}
\end{eqnarray}
where $q_{e\bot}$ is heat flow in the direction of fast ion flux and $q_{e\wedge}$ is orthogonal to that flux and $\mathbf{B}$ field. Note that, given this geometry $\left(j^{0}_{e\alpha}\right)_{\bot}$ has contributions from both $\mathbf{j}^{0}_{\alpha}$ and $\boldsymbol{\xi}_{e\alpha}$ whilst $\left(j^{0}_{e\alpha}\right)_{\wedge}$ has contributions only from $\boldsymbol{\xi}_{e\alpha}$.

\begin{figure}
\begin{center}%
\includegraphics*[width=0.95\columnwidth]{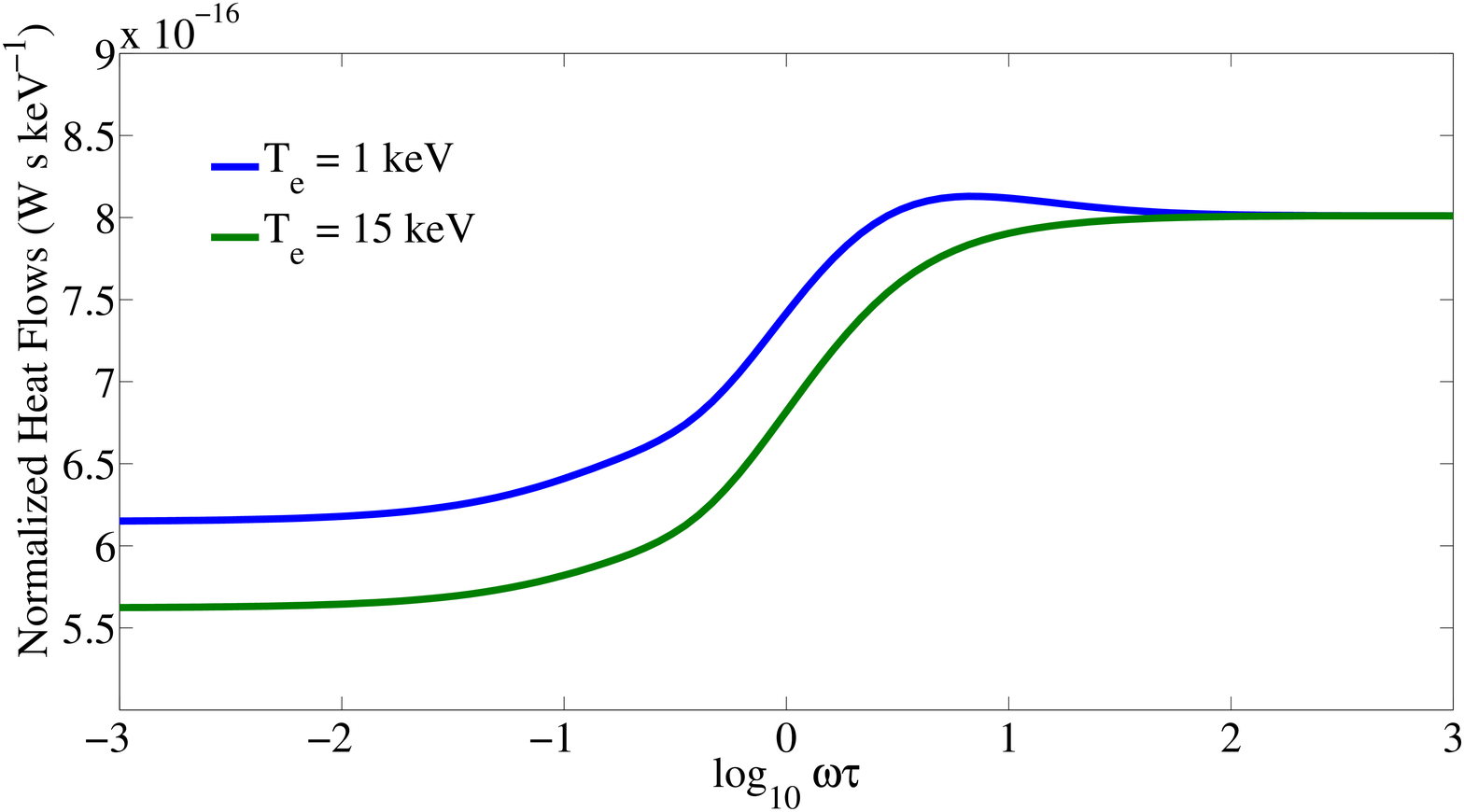}%
\hfil
\includegraphics*[width=0.95\columnwidth]{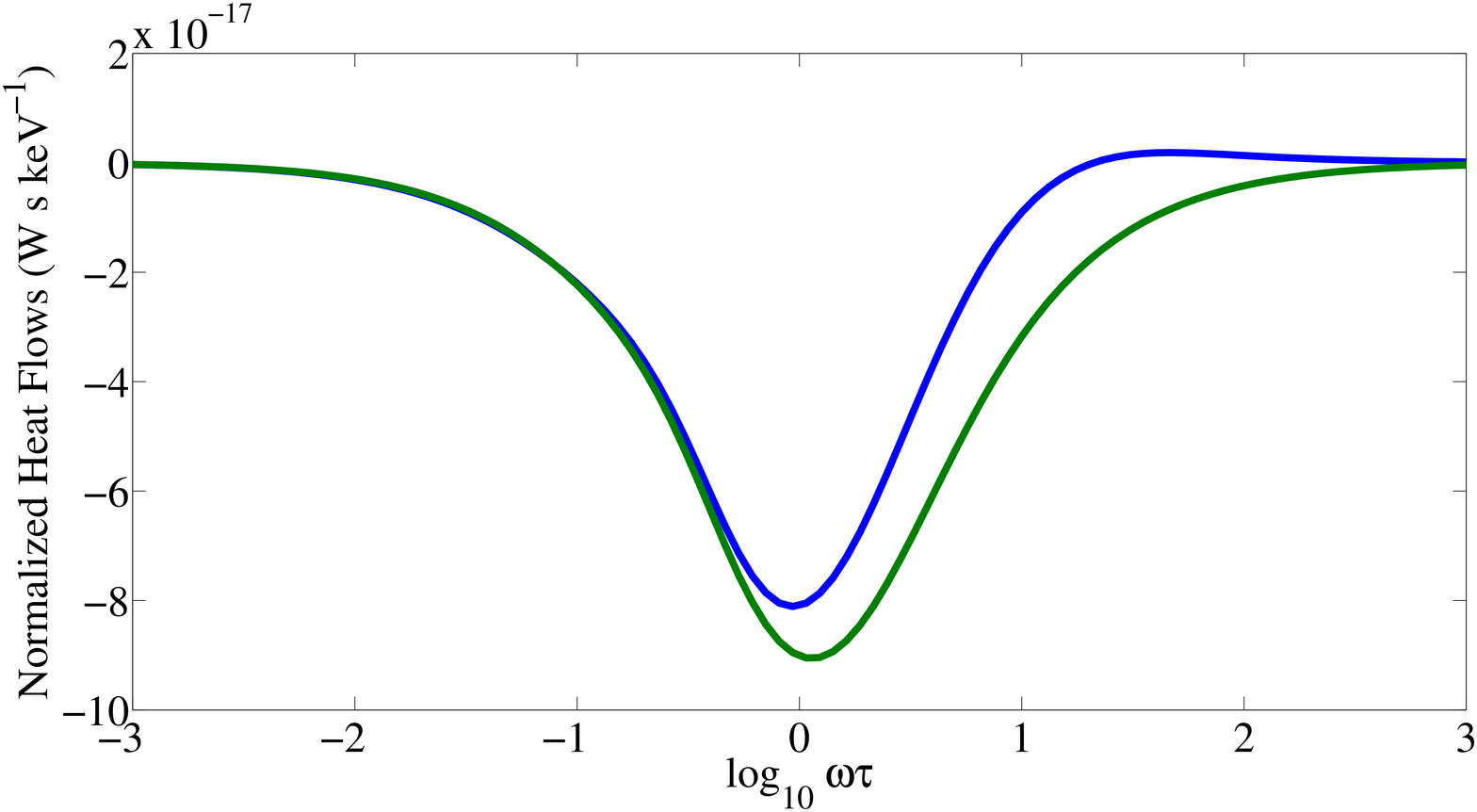}%
\vspace{-1em}
\end{center}
\caption[]{Normalized heat flows due to electron-fast ion collisions as a function of $\omega\tau$ where the fast ions are a mono-energetic population of $\alpha$ particles with $E_{\alpha} = 3.45\,MeV$. Results for two different values of $T_{e}$ are shown. The heat flow values have been normalized by the $\alpha$ flux $n_{\alpha}\langle v_{\alpha}\rangle$ and $T_{e}$ in units of $keV$. Top: The component parallel to the $\alpha$ flux, $q_{e\bot}$.  Bottom: The components orthogonal to the $\alpha$ flux and $\mathbf{B}$ field, $q_{e\wedge}$.} \label{f:6}
\end{figure}

Figure \ref{f:6} shows the variation of these components of heat flow with $\omega\tau$ for $3.45\,MeV$ $\alpha$ particles in a DT plasma. The most remarkable result is that the value of $q_{e\bot}$ increases with increasing magnetization. This is because magnetic fields reduce the values of both $\left(\boldsymbol{\xi}_{e\alpha}\right)_{\bot}$ and $\left(\boldsymbol{\zeta}_{e\alpha}\right)_{\bot}$ since the electrons become magnetically confined. However, the $\mathbf{j}_{\alpha}^{0}$ is due to the motion of $\alpha$ particles and, since these have a much larger Larmor radius than the electrons, we expect that far higher values of $\omega\tau$ than considered here are required to reduce the value of $\mathbf{j}_{\alpha}^{0}$. Therefore, $q_{e\bot}$ is dominated by $\left(\boldsymbol{\zeta}_{e\alpha}\right)_{\bot}$ in unmagnetized plasmas and at low values of $\omega\tau$ but at high values of $\omega\tau$ it is $\mathbf{j}_{\alpha}^{0}$ which dominates.

In contrast, the $q_{e\wedge}$ component is largest at intermediate values of $\omega\tau$, as we might expect from the behaviour of $\left(\boldsymbol{j}^{0}_{e\alpha}\right)_{\wedge}$ and $\left(\boldsymbol{\zeta}_{e\alpha}\right)_{\wedge}$ shown in fig. \ref{f:3}. In this case, $\mathbf{j}_{\alpha}^{0}$ (which appears in the $\left(\boldsymbol{j}_{e\alpha}\right)_{\bot}$ term) is unable to make a significant contribution to heat flow at large values of $\omega\tau$ as it is multiplied by a factor of $\beta_{\wedge}$, which decreases rapidly with increasing magnetization.

We can conclude that the heat flow induced by $\alpha$-electron collisions is not suppressed by the magnetization of the electrons. However, electron thermal conduction is reduced by several orders of magnitude at large values of $\omega\tau$. Thus, it is quite possible that in a magnetized hotspot the dominant source of heat flow will be the $\alpha$-electron collisions.

This could be of interest in a number of scenarios in which large magnetic fields are present in burning inertial fusion plasmas. These include self-generated magnetic fields\cite{Walsh_PRL2017} and schemes in which a large magnetic field is imposed on the plasma in order to suppress electron thermal conduction and reduce heat losses from the hot fuel. Examples of such schemes that are currently being investigated include indirect-drive ICF with an imposed magnetic field\cite{Perkins_POP2017,Perkins_POP2013} and magneto-inertial fusion schemes such as MagLIF.\cite{Slutz_PRL2012,Slutz_POP2010}




\section{Magnetic field effects of electron-fast ion collisions}\label{sec:5}
Combining Ohm's law \eqref{e:3.2.6} with Faraday's law $\frac{\partial \mathbf{B}}{\partial t}=-\nabla\times\mathbf{E}$ gives an induction equation for the plasma which describes how the magnetic field evolves
\begin{equation}\label{e:5.1}
\frac{\partial \mathbf{B}}{\partial t}=\nabla\times\left[\frac{\nabla P_{e}}{e n_{e}}  -\underline{\underline{\boldsymbol{\sigma}}}^{-1}\cdot\left(\mathbf{j}_{T}-\mathbf{j}_{e\alpha}^{0}\right)+\frac{1}{e}\underline{\underline{\boldsymbol{\beta}}} \cdot\nabla T_{e}\right],
\end{equation}
The approach of Braginskii and others is to separate the inverse electrical conductivity term into an electrical resistivity $\underline{\underline{\boldsymbol{\alpha}}}$ and a term proportional to both current and magnetic field, often referred to as the Hall term, as follows
\begin{equation}\label{e:5.2}
\underline{\underline{\boldsymbol{\sigma}}}^{-1}\cdot\left(\mathbf{j}_{e\alpha}^{0}-\mathbf{j}_{T}\right)=\frac{1}{en_{e}}\left[\left(\mathbf{j}_{e\alpha}^{0}-\mathbf{j}_{T}\right)\times\mathbf{B}+\frac{1}{en_{e}}\underline{\underline{\boldsymbol{\alpha}}}\cdot\left(\mathbf{j}_{e\alpha}^{0}-\mathbf{j}_{T}\right)\right].
\end{equation}
where the resistivity tensor $\underline{\underline{\boldsymbol{\alpha}}}$ obeys the following relation
\begin{equation}\label{e:5.3}
\underline{\underline{\boldsymbol{\alpha}}}\cdot\mathbf{s} = \alpha_{\|}\left(\mathbf{b}\cdot\mathbf{s}\right)\mathbf{b}+\alpha_{\bot}\mathbf{b}\times\left(\mathbf{s}\times\mathbf{b}\right)-\alpha_{\wedge}\mathbf{b}\times\mathbf{s}.
\end{equation}
Note, the minus sign on the last term on the RHS in contrast to \eqref{e:3.2.10}. The induction equation can now be written as
\begin{eqnarray}\label{e:5.4}
\frac{\partial \mathbf{B}}{\partial t}&=&\frac{1}{en_{e}}\nabla T_{e}\times\nabla n_{e}+\frac{1}{e}\nabla\times\left[\underline{\underline{\boldsymbol{\beta}}} \cdot\nabla T_{e}\right]-\nabla\times\frac{1}{en_{e}}\left[\mathbf{j}_{T}\times\mathbf{B}+\frac{1}{en_{e}}\underline{\underline{\boldsymbol{\alpha}}}\cdot\mathbf{j}_{T}\right]\nonumber\\
&&+\nabla\times\frac{1}{en_{e}}\left[\mathbf{j}_{e\alpha}^{0}\times\mathbf{B}+\frac{1}{en_{e}}\underline{\underline{\boldsymbol{\alpha}}}\cdot\mathbf{j}_{e\alpha}^{0}\right],
\end{eqnarray}
The magnetic field $\mathbf{B}$ and total current $\mathbf{j}_{T}$ terms are related by Ampere's law, $\mathbf{j}_{T}=\frac{1}{\mu_{0}}\nabla\times\mathbf{B}$, and so the terms containing $\mathbf{j}_{e\alpha}^{0}$ can be treated as truly independent, i.e. when fast ions collide with electrons and generate a current $\mathbf{j}_{e\alpha}^{0}$, the last two terms in \eqref{e:5.4} determine the effect on the magnetic field, regardless of the existing current and magnetic field in the plasma. We now examine each of those terms.

\subsection{Magnetic field advection}\label{sec:5.1}
We start with the $\mathbf{j}_{e\alpha}^{0}\times\mathbf{B}$ which is analogous to the Hall term in the conventional Ohm's law
\begin{equation}\label{e:5.1.1}
\frac{\partial \mathbf{B}}{\partial t}=\nabla\times\frac{\mathbf{j}_{e\alpha}^{0}}{en_{e}}\times\mathbf{B}
\end{equation}
This is an advection equation in which $\frac{\mathbf{j}_{e\alpha}^{0}}{en_{e}}$ represents an advection velocity.

From the results in fig. \ref{f:3} we can conclude that for $\alpha$ particles, $\left|\mathbf{j}_{e\alpha}^{0}\right| \sim \left|\mathbf{j}_{\alpha}^{0}\right|$. Therefore, we can estimate the advection velocity magnitude to be $\approx \frac{2n_{\alpha}}{n_{e}}\langle\mathbf{v}_{\alpha}\rangle$. It is notable that the direction of the advection velocity will depend on the value of $\omega\tau$, as illustrated in fig. \ref{f:3}. In the recent ICF experiments, described in \ref{sec:4.1}, the hotspot density was estimated to be $\sim 10^{5}\,kg\,m^{-3}$. Therefore, using the estimate of $n_{\alpha}$ from above, we have $n_{\alpha}/n_{e}\sim 10^{-3}$. Taking an upper limit of $10^{7}\,m\,s^{-1}$ for $\langle\mathbf{v}_{\alpha}\rangle$ we can estimate advection velocities of up to $10^{4}\,m\,s^{-1}$ in such a hotspot, assuming a magnetic field is present.

\subsection{Magnetic field generation}\label{sec:5.2}
In an unmagnetized plasma we have $\mathbf{B} = 0$ and $\mathbf{j}_{T} = 0$ and so \eqref{e:5.4} reduces to
\begin{equation}\label{e:5.2.1}
\frac{\partial \mathbf{B}}{\partial t} = \frac{1}{en_{e}}\nabla T_{e}\times\nabla n_{e} + \nabla\times\left[\frac{\alpha_{\|}}{\left(en_{e}\right)^{2}}\mathbf{j}_{e\alpha}^{0}\right].
\end{equation}
The first term on the RHS is the well-known Biermann battery term\cite{Biermann_1950} and will not be considered here. The second term arises from the electron-fast ion collisions and exists even when the total current is zero, i.e. collisions between electrons and fast ions can induce a magnetic field in an initially unmagnetized plasma.

For a plasma with $Z_{i} = 1$, the unmagnetized electrical resistivity, as given by Epperlein and Haines\cite{Epperlein_1986}, is
\begin{equation}\label{e:5.2.2}
\alpha_{\|} = 0.506\frac{m_{e}n_{e}}{\tau_{ei}}.
\end{equation}
Using this result and working in cylindrical co-ordinates, the induction equation due to an axially directed $\mathbf{j}_{e\alpha}^{0}$ becomes
\begin{equation}\label{e:5.2.4}
\frac{\partial B_{\theta}}{\partial t} = 1.65\times10^{-9}\frac{\ln\Lambda_{ei}}{T_{keV}^{\frac{3}{2}}}\left[-\frac{\partial j_{e\alpha}^{0}}{\partial r}+\frac{3}{2}\frac{j_{e\alpha}^{0}}{T_{keV}}\frac{\partial T_{keV}}{\partial r}\right]\,T\,s^{-1},
\end{equation}
where we have assumed $n_{e}\approx n_{i}$ and $T_{keV}$ is the electron temperature in units of $keV$. This equation is applied to two different scenarios. The first is ion fast ignition fusion\cite{Fernandez_NucFus2014} in which cold DT fuel is compressed to densities of $~400\,g\,cm^{-3}$ and then heated to temperatures of $>10\,keV$ using an ion beam delivering a power density of $\approx 10^{22}\,W\,cm^{-3}$ in a time of $\approx20\,ps$. Numerical studies\cite{Honrubia_POP2015} have shown that carbon ion beams with an average particle energy of $200\,MeV$ are an efficient way to deliver this power. Such a beam corresponds to a fast ion current of $\mathbf{j}_{\alpha}^{0} = 3.4\times10^{17}\,A\,m^{-3}$. Using the results from fig. \ref{f:2} we can estimate $\mathbf{j}_{e\alpha}^{0} \approx -5\mathbf{j}_{\alpha}^{0}$. Assuming an ion beam radius of $10\,\mu m$ and a spatially uniform electron temperature in the range $T_{e}=1-10\,keV$, \eqref{e:5.2.4} results field generation rates of $10^{13}-10^{15}\,T\,s^{-1}$.

The second scenario to which we apply \eqref{e:5.2.4} is a flux of $\alpha$ particles moving orthogonal to an electron temperature gradient. In section \ref{sec:4.1} we estimated that $\alpha$ particle fluxes of $10^{34}\,m^{-2}\,s^{-1}$ may exist in current ICF experiment. This corresponds to $\mathbf{j}_{e\alpha}^{0} \approx 6\times 10^{15}\,A\,m^{-3}$. We assume that this current is spatially uniform and flowing orthogonal to a temperature gradient of $1\,keV\,\mu m^{-1}$. From \eqref{e:5.2.4} we can estimate that this will give rise to a field generation rate of $6\times10^{13}T_{keV}^{-\frac{5}{2}}\,T\,s^{-1}$ as the $\alpha$ particles move across the temperature gradient.

Both these examples demonstrate that significant field generation can occur when a flux of fast ions is undergoing collisions with a thermal plasma. Modelling the evolution and effects of such a field will require solving a time-dependent set of fluid-kinetic equations that will be the subject of future work.





\section{Modifications of transport coefficients due to the isotropic fast ion component}\label{sec:6}
We have focussed so far on the effects of the anisotropic component of the fast ion population, $\mathbf{F_{1}}$. The terms $\mathbf{j}_{e\alpha}^{0}$ and $\boldsymbol{\zeta}_{e\alpha}$, which we have introduced into the classical electron transport equations, result from this anisotropy and equal zero if the fast ion population is isotropic ($\mathbf{F_{1}} = 0$). The isotropic component of the fast ion distribution, $F_{0}$, appears on the LHS of \eqref{e:3.1.3} and, therefore, cannot drive transport. However, the $F_{0}$ component can affect the perturbation of the electrons, $\mathbf{f_{1}}$, for a given driving term on the RHS of \eqref{e:3.1.3}.

By examining the functions in which $F_{0}$ appears ($g_{1-3}\left(v\right)$, listed in \eqref{e:a2.7}-\eqref{e:a2.12}) we can see that there is an analogous term involving $f_{0}$ for each $F_{0}$ term. If we assume $n_{\alpha}\ll n_{e}$, which we have done so far in this work, then the $F_{0}$ terms will be of minor importance in the LHS of \eqref{e:3.1.3}. This result has allowed us to use conventional values (i.e. values calculated when no fast ions are present\cite{Epperlein_1986}) for the electric conductivity, thermoelectric, energy conductivity and thermal diffusion tensors in \eqref{e:3.2.1} and \eqref{e:3.2.2}.

\begin{figure}
\begin{center}%
\includegraphics*[width=0.95\columnwidth]{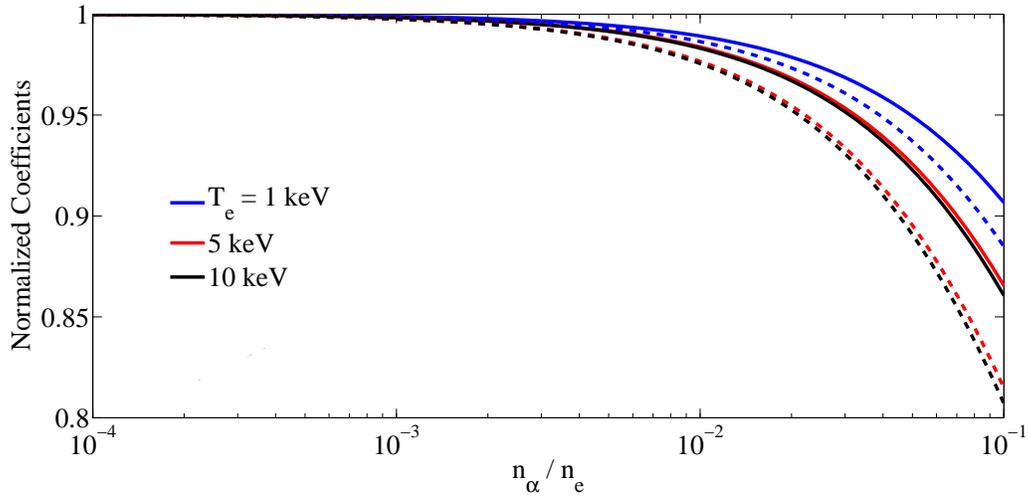}%
\vspace{-1em}
\end{center}
\caption[]{Values of the energy conductivity, $\mu_{\|}$, and thermal diffusion, $K_{\|}$, as a function of the number density ratio of mono-energetic $3.45\,MeV$ $\alpha$ particles to electrons for three values of electron temperature. Both $\mu_{\|}$ and $K_{\|}$ are normalized by their values when no $\alpha$ particles are present. Full lines represent $K_{\|}$ and dashed lines represent $\mu_{\|}$.} \label{f:8}
\end{figure}

We investigate the validity of this approach by calculating values of energy conductivity, $\mu_{\|}$, and thermal diffusion, $K_{\|}$, for an unmagnetized plasma containing an isotropic population of $3.45\,MeV$ $\alpha$ particles. Figure \ref{f:8} contains results for these calculations as a function of the ratio of $\alpha$ particle density to electron density. It shows that for $n_{\alpha}/n_{e}<\sim10^{-3}$ the calculated values of $\mu_{\|}$ and $K_{\|}$ deviate by less than $1\%$ from their conventional values. However, for $n_{\alpha}/n_{e}>\sim10^{-2}$ the values $\mu_{\|}$ and $K_{\|}$ begin to deviate significantly. Therefore, we can conclude that when $n_{\alpha}\ll n_{e}$, the isotropic component of the fast ion distribution, $F_{0}$, will not affect the electron transport. This assumption was implicit in our work in sections \ref{sec:3}-\ref{sec:5}. The value $n_{\alpha}/n_{e}$ at which this assumption breaks down will depend on $v_{a}/v_{Te}$. When it breaks down it will be necessary to consider the effects of isotropic fast ions on the electron transport, not just the flux component.

\section{Conclusions}\label{sec:7}
In this work we have demonstrated that Coulomb collisions between electrons and a population of fast ions can perturb the electron distribution function from Maxwellian even when the density of fast ions is much less than the electron density. Anisotropy of the fast ion distribution drives the perturbation. These perturbations directly affect the transport of charge and energy by the electrons. We have derived a set of classical electron transport equations that include the effects of the these perturbations. These equations are an Ohm's law, which includes an additional current term due to electron-fast ion collisions, and a heat flow equation, which includes two additional terms.

These new transport equations have been examined and a number of interesting effects have been highlighted. First, it was shown that a flux of $\alpha$ particles in a DT plasma will induce an electron heat flow in the same direction as the $\alpha$ flux. Notably, it was shown that even when the electrons are magnetized, the flux of $\alpha$ particles across the field lines will induce a heat flow across the field lines that cannot be suppressed by the magnetic field. Secondly, it was shown that $\alpha$ particles will advect magnetic field in the plasma with a velocity that is proportional to the drift velocity of the $\alpha$ population and the ratio of $\alpha$ particle density to electron density. Thirdly, it was shown that magnetic field can be generated by the new resistivity term in Ohm's law arising from the current induced by electron-fast ion collisions.

The transport equations are suitable for inclusion in integrated MHD codes. This will be the subject of future work and will help to elucidate the effects that we have outlined here.


\section{Appendix}
\subsection{Expressions for collision terms in the electron $\mathbf{f_{1}}$ equations}\label{app:2}
In this section we list terms for the linearized collision operators used in the Fokker-Planck model. We first define the expression
\begin{equation}\label{e:a2.1}
Y_{ab} = \frac{1}{4\pi}\left(\frac{q_{a}q_{b}}{\varepsilon_{0}m_{a}}\right)^{2}\ln\Lambda_{ab},
\end{equation}
where $q_{a}$ and $m_{a}$ is the charge and mass of particles of species $a$ and $\ln\Lambda_{ab}$ is the Coulomb logarithm.

The Fokker-Planck equation can be used to model the Coulomb collisions between distributions of charged particles. We have expanded the electron distribution function in a series of Cartesian tensors \eqref{e:2.2} and used the $0$th and $1$st Cartesian tensor for the $\alpha$ distribution \eqref{e:3.1.1}. These functions can be inserted into a general form of the Fokker-Planck equation and the first angular moment can be taken to find the collision terms necessary for solving \eqref{e:2.3}. This procedure has been carried out by Shkarofsky and co-workers and we use their result (equation $7-75$ of Shkarofsky\cite{Shkarofsky_1966}) for electron-electron collisions. This is
\begin{eqnarray}\label{e:a2.2}
\mathbf{C}_{ee1}&=&\frac{Y_{ee}}{3v}\left(I_{2}^{e0}+J_{-1}^{e0}\right)\frac{\partial^{2} \mathbf{f_{1}}}{\partial v^{2}}\nonumber\\
&&+\frac{Y_{ee}}{3v^{2}}\left(3I_{0}^{e0}-I_{2}^{e0}+2J_{-1}^{e0}\right)\frac{\partial \mathbf{f_{1}}}{\partial v}\nonumber\\
&&+\frac{Y_{ee}}{3v^{3}}\left(-3I_{0}^{e0}+I_{2}^{e0}-2J_{-1}^{e0}\right)+8\pi Y_{ee} f_{0}\mathbf{f_{1}}\nonumber\\
&&+\frac{Y_{ee}}{5v}\left(I_{3}^{e1}+J_{-2}^{e1}\right)\frac{\partial^{2} f_{0}}{\partial v^{2}}\nonumber\\
&&+\frac{Y_{ee}}{15v^{2}}\left(-3I_{3}^{e1}+2J_{-2}^{e1}+5I_{1}^{e1}\right)\frac{\partial f_{0}}{\partial v},
\end{eqnarray}
where the $I$ and $J$ functions are defined as
\begin{eqnarray}
I_{j}^{e i} &=& \frac{4\pi}{v^{j}}\int_{0}^{v}V^{j+2}f_{i}dV,\label{e:a2.3}\\
J_{j}^{e i} &=& \frac{4\pi}{v^{j}}\int_{v}^{\infty}V^{j+2}f_{i}dV,\label{e:a2.4}
\end{eqnarray}

Following a similar approach for electron-fast ion collisions gives
\begin{eqnarray}\label{e:a2.5}
\mathbf{C}_{e\alpha 1} = \mathbf{C}_{e\alpha1}^{01}+\mathbf{C}_{e\alpha1}^{10} &=&\frac{Y_{e\alpha}}{3v}\left(I_{2}^{\alpha0}+J_{-1}^{\alpha0}\right)\frac{\partial^{2} \mathbf{f_{1}}}{\partial v^{2}}\nonumber\\
&&+\frac{Y_{e\alpha}}{3v^{2}}\left(3\frac{m_{e}}{m_{\alpha}}I_{0}^{\alpha0}-I_{2}^{\alpha0}+2J_{-1}^{\alpha0}\right)\frac{\partial \mathbf{f_{1}}}{\partial v}\nonumber\\
&&+\frac{Y_{e\alpha}}{3v^{3}}\left(-3I_{0}^{\alpha0}+I_{2}^{\alpha0}-2J_{-1}^{\alpha0}\right)+4\pi Y_{e\alpha}\frac{m_{e}}{m_{\alpha}}\left(F_{0}\mathbf{f_{1}}+f_{0}\mathbf{F_{1}}\right)\nonumber\\
&&+\frac{Y_{e\alpha}}{5v}\left(I_{3}^{\alpha1}+J_{-2}^{\alpha1}\right)\frac{\partial^{2} f_{0}}{\partial v^{2}}\nonumber\\
&&+\frac{Y_{e\alpha}}{15v^{2}}\left(-3I_{3}^{\alpha1}+\left(7-5\frac{m_{e}}{m_{\alpha}}\right)J_{-2}^{\alpha1}+\left(-5+10\frac{m_{e}}{m_{\alpha}}\right)I_{1}^{\alpha1}\right)\frac{\partial f_{0}}{\partial v},
\end{eqnarray}
with $I$ and $J$ functions
\begin{eqnarray}
I_{j}^{\alpha i} &=& \frac{4\pi}{v^{j}}\int_{0}^{v}V^{j+2}F_{i}dV,\label{e:a2.5}\\
J_{j}^{\alpha i} &=& \frac{4\pi}{v^{j}}\int_{v}^{\infty}V^{j+2}F_{i}dV.\label{e:a2.6}
\end{eqnarray}

Including \eqref{e:a2.2} and \eqref{e:a2.3} in the electron $\mathbf{f_{1}}$ equation, \eqref{e:3.1.2}, and carrying out some algebra results in \eqref{e:3.1.3} where the functions $g_{1}$-$g_{6}$ are defined by
\begin{eqnarray}
g_{1}\left(v\right)&=& \tau_{ei}\frac{Y_{ee}}{3v}\left(I_{2}^{e0}+J_{-1}^{e0}\right)+\tau_{ei}\frac{Y_{e\alpha}}{3v}\left(I_{2}^{\alpha0}+J_{-1}^{\alpha0}\right),\label{e:a2.7}\\
g_{2}\left(v\right)&=& \tau_{ei}\frac{Y_{ee}}{3v^{2}}\left(3I_{0}^{e0}-I_{2}^{e0}+2J_{-1}^{e0}\right)+\tau_{ei}\frac{Y_{e\alpha}}{3v^{2}}\left(3\frac{m_{e}}{m_{\alpha}}I_{0}^{\alpha0}-I_{2}^{\alpha0}+2J_{-1}^{\alpha0}\right),\label{e:a2.8}\\
g_{3}\left(v\right)&=& -\tau_{ei} Y_{ei}n_{i}\frac{1}{v^{3}}+8\pi \tau_{ei} Y_{ee} f_{0}+\frac{\tau_{ei} Y_{ee}}{3v^{3}}\left(-3I_{0}^{e0}+I_{2}^{e0}-2J_{-1}^{e0}\right)\nonumber\\
&&+4\pi \tau_{ei} Y_{e\alpha}\frac{m_{e}}{m_{\alpha}}F_{0}+\tau_{ei}\frac{Y_{e\alpha}}{3v^{3}}\left(-3I_{0}^{\alpha0}+I_{2}^{\alpha0}-2J_{-1}^{\alpha0}\right),\label{e:a2.9}\\
g_{4}\left(v\right)&=& 16\pi\frac{\tau_{ei} Y_{ee}}{5v_{Te}^{4}}\frac{1}{v^{2}}f_{0},\label{e:a2.10}\\
g_{5}\left(v\right)&=& -8\pi\frac{\tau_{ei} Y_{ee}}{3v_{Te}^{2}}\frac{1}{v^{2}}f_{0},\label{e:a2.11}\\
g_{6}\left(v\right)&=& 8\pi\frac{\tau_{ei} Y_{ee}}{5v_{Te}^{2}}\left(2\frac{v^{2}}{v_{Te}^{2}}-\frac{5}{3}\right)vf_{0}.\label{e:a2.12}
\end{eqnarray}
We emphasise that these are scalar functions, depending only on the magnitude of the velocity, $v$.

\subsection{Solving the linear system of equations}\label{app:3}
We wish to solve the following linear integro-differential equation for the function $\mathbf{f_{1}}\left(v\right)$
\begin{eqnarray}\label{e:a3.1}
&&g_{1}\left(v\right)\frac{\partial^{2} \mathbf{f_{1}}}{\partial v^{2}}+g_{2}\left(v\right)\frac{\partial \mathbf{f_{1}}}{\partial v}+g_{3}\left(v\right)\mathbf{f_{1}}+g_{4}\left(v\right)\int_{0}^{v}v^{5}\mathbf{f_{1}}dv\nonumber\\
&&+g_{5}\left(v\right)\int_{0}^{v}v^{3}\mathbf{f_{1}}dv+g_{6}\left(v\right)\int_{v}^{\infty}\mathbf{f_{1}}dv+\omega\tau_{ei}\mathbf{b}\times\mathbf{f_{1}}=\mathbf{g_{7}}\left(v\right),
\end{eqnarray}
where the terms $g_{1-7}$ depend only on $v$ and parameters such as $T_{e}$, $n_{e}$, $F_{0}$, $\mathbf{F_{1}}$, etc. We follow a similar method to that of Epperlein and Haines.\cite{Epperlein_1986} Defining the linear operators
\begin{eqnarray}\label{e:a3.2}
G &=& g_{1}\frac{\partial^{2} }{\partial v^{2}}+g_{2}\frac{\partial }{\partial v}+g_{3}+g_{4}\int_{0}^{v}dvv^{5}+g_{5}\int_{0}^{v}dv v^{3}+g_{6}\int_{v}^{\infty}dv,\\
\boldsymbol{\Omega} &=& \omega\tau_{ei}\mathbf{b}
\end{eqnarray}
allows us to write \eqref{e:a3.1} as
\begin{equation}\label{e:a3.3}
\left(G+\boldsymbol{\Omega}\times\right)\mathbf{f_{1}} = \mathbf{g_{7}}.
\end{equation}
Operating on this equation with $G$ and $-\boldsymbol{\Omega}\times$ and summing the results gives
\begin{equation}\label{e:a3.4}
\left(GG+\Omega^2\right)\mathbf{f_{1}}-\left(\boldsymbol{\Omega}\cdot\mathbf{f_{1}}\right)\boldsymbol{\Omega} = \left(G-\boldsymbol{\Omega}\times\right)\mathbf{g_{7}}.
\end{equation}
Without loss of generality, we assume a Cartesian co-ordinate system with $\mathbf{b}$ in the $+z$ direction and so, from \eqref{e:a3.3} and \eqref{e:a3.4}, we get
\begin{eqnarray}
\left(GG+\Omega^2\right)f_{1x} &=& \left(G+\Omega\right)g_{7x},\label{e:a3.5}\\
\left(GG+\Omega^2\right)f_{1y} &=& \left(G-\Omega\right)g_{7y},\label{e:a3.6}\\
Gf_{1z} &=& g_{7z}.\label{e:a3.7}
\end{eqnarray}
We wish to solve these equations for $\mathbf{f_{1}}=\left[f_{1x},f_{1y},f_{1z}\right]$ for a given $\mathbf{g_{7}}=\left[g_{7x},g_{7y},g_{7z}\right]$, where $f_{1z}$ is the component parallel to the magnetic field and $f_{1x}$ and $f_{1y}$ are orthogonal. We assume the following boundary conditions for $\mathbf{f_{1}}$
\begin{eqnarray}
\mathbf{f_{1}}\left(\infty\right)&=&0,\\
\frac{\partial}{\partial v}\mathbf{f_{1}}\left(0\right)&=&0,
\end{eqnarray}
The equations \eqref{e:a3.5}-\eqref{e:a3.7} are solved by finite differencing on the interval $v\in\left[0,v_{max}\right]$. We choose $N$ (where $N$ is odd) evenly spaced values of $v$ on this interval such that
\begin{eqnarray}
v_{j}&=&\left(j-1\right)h,\\
h&=&\frac{v_{max}}{N-1},\\
j&=&1,2,\ldots,N,
\end{eqnarray}
For each equation of \eqref{e:a3.5}-\eqref{e:a3.7}, we wish to find the matrix of values $\mathbf{f_{1}}^{1},\mathbf{f_{1}}^{2},\ldots,\mathbf{f_{1}}^{j},\ldots,\mathbf{f_{1}}^{N}$ corresponding to the velocities $v_{1},v_{2},\ldots,v_{j},\ldots,v_{N}$. The discretized boundary conditions are
\begin{eqnarray}
\mathbf{f_{1}}^{1}&=&\mathbf{f_{1}}^{2},\\
\mathbf{f_{1}}^{N}&=&0,
\end{eqnarray}
and so we only need to calculate values for $\mathbf{f_{1}}^{2},\ldots,\mathbf{f_{1}}^{j},\ldots,\mathbf{f_{1}}^{N-1}$. The values $\mathbf{g_{7}}^{2},\ldots,\mathbf{g_{7}}^{j},\ldots,\mathbf{g_{7}}^{N-1}$ can be calculated for the corresponding velocities whilst the discretized $\Omega^2$ and $\Omega$ terms are $\Omega^2\underline{\underline{\mathbf{I}}}$ and $\Omega\underline{\underline{\mathbf{I}}}$, respectively, where $\underline{\underline{\mathbf{I}}}$ is the $N-2\times N-2$ identity matrix. Finally, the discretized operator $G$ is $N-2\times N-2$ matrix whose entries are obtained from the discretized differential and integral terms. We use five-point differencing for the derivatives as follows
\begin{equation}
g_{1}^{j}\frac{\partial^{2} \mathbf{f_{1}}^{j}}{\partial v^{2}} = \begin{cases}
g_{1}^{j}\left(\frac{-\mathbf{f_{1}}^{j+2}+16\mathbf{f_{1}}^{j+1}-30\mathbf{f_{1}}^{j}+16\mathbf{f_{1}}^{j-1}-\mathbf{f_{1}}^{j-2}}{12h^{2}}\right),\quad\,2<j<N-1,\nonumber\\
g_{1}^{j}\left(\frac{\mathbf{f_{1}}^{j-1}-2\mathbf{f_{1}}^{j}+\mathbf{f_{1}}^{j+1}}{h^{2}}\right),\quad\,j=2,j=N-1,\nonumber\end{cases}
\end{equation}
and
\begin{equation}
g_{2}^{j}\frac{\partial \mathbf{f_{1}}^{j}}{\partial v} = \begin{cases}
g_{2}^{j}\left(\frac{-\mathbf{f_{1}}^{j+2}+8\mathbf{f_{1}}^{j+1}-8\mathbf{f_{1}}^{j-1}+\mathbf{f_{1}}^{j-2}}{12h}\right),\quad\,2<j<N-1,\nonumber\\
g_{2}^{j}\left(\frac{\mathbf{f_{1}}^{j+1}-\mathbf{f_{1}}^{j-1}}{2h}\right),\quad\,j=2,j=N-1.\nonumber\end{cases}
\end{equation}
We use Simpson's rule for integration as follows
\begin{equation}
g_{4}^{j}\int_{0}^{v_{j}}v^{5}\mathbf{f_{1}}^{j}dv=\begin{cases}
g_{4}^{j}\frac{h}{3}\left[v_{1}^{5}\mathbf{f_{1}}^{1}+4\left(v_{2}^{5}\mathbf{f_{1}}^{2}+v_{4}^{5}\mathbf{f_{1}}^{4}+\ldots+v_{j-1}^{5}\mathbf{f_{1}}^{j-1}\right)\right.\nonumber\\
\left.\qquad+2\left(v_{3}^{5}\mathbf{f_{1}}^{3}+v_{5}^{5}\mathbf{f_{1}}^{5}+\ldots+v_{j-2}^{5}\mathbf{f_{1}}^{j-2}\right)+v_{j}^{5}\mathbf{f_{1}}^{j}\right],\quad\,j=3,5,\ldots,N-2,\nonumber\\
g_{4}^{j}\frac{h}{2}\left[v_{1}^{5}\mathbf{f_{1}}^{1}+v_{2}^{5}\mathbf{f_{1}}^{2}\right]+g_{4}^{j}\frac{h}{3}\left[v_{2}^{5}\mathbf{f_{1}}^{2}+4\left(v_{3}^{5}\mathbf{f_{1}}^{3}+v_{5}^{5}\mathbf{f_{1}}^{5}+\ldots+v_{j-1}^{5}\mathbf{f_{1}}^{j-1}\right)\right.\nonumber\\
\left.\qquad+2\left(v_{4}^{5}\mathbf{f_{1}}^{4}+v_{6}^{5}\mathbf{f_{1}}^{6}+\ldots+v_{j-2}^{5}\mathbf{f_{1}}^{j-2}\right)+v_{j}^{5}\mathbf{f_{1}}^{j}\right],\quad\,j=2,4,\ldots,N-1.\nonumber
\end{cases}
\end{equation}
The same scheme can be used for the $g_{5}\left(v\right)\int_{0}^{v}v^{3}\mathbf{f_{1}}dv$ term. The final integration term is expressed as
\begin{equation}
g_{6}^{j}\int_{v_{j}}^{\infty}\mathbf{f_{1}}^{j}dv=\begin{cases}
g_{6}^{j}\frac{h}{3}\left[\mathbf{f_{1}}^{j}+4\left(\mathbf{f_{1}}^{j+1}+\mathbf{f_{1}}^{j+3}+\ldots+\mathbf{f_{1}}^{N-1}\right)\right.\nonumber\\
\left.\qquad+2\left(\mathbf{f_{1}}^{j+2}+\mathbf{f_{1}}^{j+4}+\ldots+\mathbf{f_{1}}^{N-2}\right)+\mathbf{f_{1}}^{N}\right],\quad\,j=3,5,\ldots,N-2,\nonumber\\
g_{6}^{j}\frac{h}{2}\left[\mathbf{f_{1}}^{N-1}+\mathbf{f_{1}}^{N}\right]+g_{6}^{j}\frac{h}{3}\left[\mathbf{f_{1}}^{j}+4\left(\mathbf{f_{1}}^{j+1}+\mathbf{f_{1}}^{j+3}+\ldots+\mathbf{f_{1}}^{N-2}\right)\right.\nonumber\\
\left.\qquad+2\left(\mathbf{f_{1}}^{j+2}+\mathbf{f_{1}}^{j+4}+\ldots+\mathbf{f_{1}}^{N-3}\right)+\mathbf{f_{1}}^{N-1}\right],\quad\,j=2,4,\ldots,N-1.\nonumber
\end{cases}
\end{equation}
The discretized operator $GG$ is obtained from matrix multiplication of $G$. The linear systems of equations obtained from the discretization are then solved using LU decomposition.

\subsection{Parameterization for fluxes of $\alpha$ particles}\label{app:4}
In this appendix we give a polynomial fit for the values of $\mathbf{j}^{0}_{e\alpha}$ and $\boldsymbol{\zeta}_{e\alpha}$ for $\alpha$ particles in a DT plasma as a function of electron hall parameter, $\omega\tau$, and the ratio of $\alpha$ particle velocity to electron thermal velocity, $v_{\alpha}/v_{Te}$. The data to which the fits are applied is shown in fig. \ref{f:3}. Values for the Coulomb Logarithm used to calculate this data are taken from the NRL Plasma Formulary.\cite{NRL_2013}

The components of $\mathbf{j}^{0}_{e\alpha}$ and $\boldsymbol{\zeta}_{e\alpha}$ are expressed as
\begin{eqnarray}
\left(\boldsymbol{j}^{0}_{e\alpha}\right)_{\bot} &=& Z_{\alpha}en_{\alpha}\langle v_{\alpha}\rangle j^{\nu}_{\bot},\\
\left(\boldsymbol{j}^{0}_{e\alpha}\right)_{\wedge} &=& Z_{\alpha}en_{\alpha}\langle v_{\alpha}\rangle j^{\nu}_{\wedge},\\
\left(\boldsymbol{\zeta}_{e\alpha}\right)_{\bot} &=& n_{\alpha}\langle v_{\alpha}\rangle T_{e}\zeta^{\nu}_{\bot},\\
\left(\boldsymbol{\zeta}_{e\alpha}\right)_{\wedge} &=& n_{\alpha}\langle v_{\alpha}\rangle T_{e}\zeta^{\nu}_{\wedge}.
\end{eqnarray}
Here, the parameters $j^{\nu}_{\bot}$, $j^{\nu}_{\wedge}$, $\zeta^{\nu}_{\bot}$ and $\zeta^{\nu}_{\wedge}$ are functions of $\omega\tau$ and $v_{\alpha}/v_{Te}$. They are given by the following polynomial fits
\begin{eqnarray}
j^{\nu}_{\bot} &=& \frac{p_{1}x^{3}+p_{2}x^{2}+p_{3}x+p_{4}}{x^{4}+q_{1}x^{3}+q_{2}x^{2}+q_{3}x+q_{4}},\\
j^{\nu}_{\wedge} &=& \frac{r_{1}x^{2}+r_{2}x+r_{3}}{x^{4}+s_{1}x^{3}+s_{2}x^{2}+s_{3}x+s_{4}},\\
\zeta^{\nu}_{\bot} &=& \frac{p'_{1}x^{3}+p'_{2}x^{2}+p'_{3}x+p'_{4}}{x^{4}+q'_{1}x^{3}+q'_{2}x^{2}+q'_{3}x+q'_{4}},\\
\zeta^{\nu}_{\wedge} &=& \frac{r'_{1}x^{2}+r'_{2}x+r'_{3}}{x^{4}+s'_{1}x^{3}+s'_{2}x^{2}+s'_{3}x+s'_{4}},
\end{eqnarray}
where $x=\log_{10}\omega\tau$. The fits are valid in the region $-3\leq x \leq 3$. The coefficients are given in tables \ref{t:1}-\ref{t:4} for these for various values of $v_{\alpha}/v_{Te}$. Also listed are the values of the parameters at $\omega\tau = 0$. 

\begin{table}\centering
\caption[My table caption]{Coefficient values for $j^{\nu}_{\bot}$.}\label{t:1}
\begin{footnotesize}
\begin{tabular}{l l l l l l l l l l l l l l l l l l l}
\hline
\hline
$v_{\alpha}/v_{Te}$ & & $\omega\tau = 0$ & & $p_{1}$ & & $p_{2}$ & & $p_{3}$ & & $p_{4}$ & & $q_{1}$ & & $q_{2}$ & & $q_{3}$ & & $q_{4}$\\
\hline
$1$ & & $-1.42$ & & $9.72$ & & $7.68$ & & $41.49$ & & $9.40$ & & $2.59$ & & $26.80$ & & $18.53$ & & $19.55$ \\
$0.5$ & & $-1.39$ & & $8.71$ & & $9.15$ & & $36.13$ & & $6.13$ & & $2.82$ & & $23.32$ & & $16.81$ & & $18.25$\\
$0.33$ & & $-1.21$ & & $8.17$ & & $9.63$ & & $33.78$ & & $6.63$ & & $2.45$ & & $22.75$ & & $16.32$ & & $18.11$\\
$0.25$ & & $-1.13$ & & $7.98$ & & $9.80$ & & $33.07$ & & $6.92$ & & $2.27$ & & $22.68$ & & $16.21$ & & $18.19$\\
$0.1$ & & $-1.02$ & & $7.74$ & & $10.16$ & & $32.96$ & & $7.73$ & & $2.00$ & & $23.10$ & & $16.31$ & & $18.77$\\
$0.025$ & & $-0.99$ & & $7.70$ & & $10.25$ & & $32.98$ & & $7.90$ & & $1.94$ & & $23.20$ & & $16.34$ & & $18.90$\\
\hline
\hline
\end{tabular}
\end{footnotesize}
\end{table}

\begin{table}\centering
\caption[My table caption]{Coefficient values for $j^{\nu}_{\wedge}$.}\label{t:2}
\begin{footnotesize}
\begin{tabular}{l l l l l l l l l l l l l l l l l}
\hline
\hline
$v_{\alpha}/v_{Te}$ & & $\omega\tau = 0$ & & $r_{1}$ & & $r_{2}$ & & $r_{3}$ & & $s_{1}$ & & $s_{2}$ & & $s_{3}$ & & $s_{4}$\\
\hline
$1$ & & $0$ & & $0.1145$ & & $0.1201$ & & $-1.074$ & & $0.993$ & & $2.213$ & & $1.252$ & & $1.170$\\
$0.5$ & & $0$ & & $0.1509$ & & $0.0895$ & & $-1.410$ & & $0.523$ & & $2.314$ & & $1.382$ & & $1.559$\\
$0.33$ & & $0$ & & $0.1447$ & & $0.0660$ & & $-1.377$ & & $0.465$ & & $2.406$ & & $1.456$ & & $1.659$\\
$0.25$ & & $0$ & & $0.1396$ & & $0.0527$ & & $-1.354$ & & $0.457$ & & $2.449$ & & $1.488$ & & $1.693$\\
$0.1$ & & $0$ & & $0.1307$ & & $0.0450$ & & $-1.283$ & & $0.461$ & & $2.459$ & & $1.496$ & & $1.699$\\
$0.025$ & & $0$ & & $0.1293$ & & $0.0445$ & & $-1.269$ & & $0.461$ & & $2.459$ & & $1.496$ & & $1.698$\\
\hline
\hline
\end{tabular}
\end{footnotesize}
\end{table}

\begin{table}\centering
\caption[My table caption]{Coefficient values for $\zeta^{\nu}_{\bot}$.}\label{t:3}
\begin{footnotesize}
\begin{tabular}{l l l l l l l l l l l l l l l l l l l}
\hline
\hline
$v_{\alpha}/v_{Te}$ & & $\omega\tau = 0$ & & $p'_{1}$ & & $p'_{2}$ & & $p'_{3}$ & & $p'_{4}$ & & $q'_{1}$ & & $q'_{2}$ & & $q'_{3}$ & & $q'_{4}$\\
\hline
$1$ & & $14.10$ & & $-14.06$ & & $61.74$ & & $-81.46$ & & $35.16$	& & $6.84$	& & $26.99$	& & $23.01$	& & $16.38$ \\
$0.5$ & & $12.30$ & & $-12.82$ & & $56.77$ & & $-74.79$ & & $31.06$	& & $6.81$	& & $27.34$	& & $22.94$	& & $16.59$\\
$0.33$ & & $11.14$ & & $-11.68$ & & $51.80$ & & $-68.22$ & & $28.16$ & & $6.81$ & & $27.39$ & & $22.93$ & & $16.62$\\
$0.25$ & & $10.68$ & & $-11.22$ & & $49.80$ & & $-65.57$ & & $27.01$ & & $6.81$ & & $27.41$ & & $22.93$ & & $16.63$\\
$0.1$ & & $10.08$ & & $-10.60$ & & $47.02$ & & $-61.91$ & & $25.49$ & & $6.81$ & & $27.41$ & & $22.93$ & & $16.63$\\
$0.025$ & & $9.96$ & & $-10.48$ & & $46.50$ & & $-61.23$ & & $25.21$ & & $6.81$ & & $27.41$ & & $22.93$ & & $16.63$\\
\hline
\hline
\end{tabular}
\end{footnotesize}
\end{table}

\begin{table}\centering
\caption[My table caption]{Coefficient values for $\zeta^{\nu}_{\wedge}$.}\label{t:4}
\begin{footnotesize}
\begin{tabular}{l l l l l l l l l l l l l l l l l}
\hline
\hline
$v_{\alpha}/v_{Te}$ & & $\omega\tau = 0$ & & $r'_{1}$ & & $r'_{2}$ & & $r'_{3}$ & & $s'_{1}$ & & $s'_{2}$ & & $s'_{3}$ & & $s'_{4}$\\
\hline
$1$ & & $0$ & & $-0.6999$ & & $-0.5442$ & & $6.606$ & & $1.552$ & & $3.133$ & & $2.013$ & & $1.356$\\
$0.5$ & & $0$ & & $-0.5771$ & & $-0.4996$ & & $5.471$	& & $1.531$ & & $2.983$ & & $1.868$ & & $1.274$\\
$0.33$ & & $0$ & & $-0.519$ & & $-0.4577$ & & $4.920$ & & $1.528$ & & $2.962$ & & $1.849$ & & $1.262$\\
$0.25$ & & $0$ & & $-0.4964$ & & $-0.4404$ & & $4.706$ & & $1.527$ & & $2.955$ & & $1.843$ & & $1.259$\\
$0.1$ & & $0$ & & $-0.4680$ & & $-0.4157$ & & $4.437$ & & $1.526$ & & $2.953$ & & $1.841$ & & $1.258$\\
$0.025$ & & $0$ & & $-0.4627$ & & $-0.4110$ & & $4.387$ & & $1.526$ & & $2.953$ & & $1.841$ & & $1.258$\\
\hline
\hline
\end{tabular}
\end{footnotesize}
\end{table}

\begin{acknowledgments}
BA would like to acknowledge Daniel Sinars and Kyle Peterson for hosting a visit to Sandia National Laboratories where part of this work was carried out and Dominic Hill at Imperial College London for informative discussions.
\end{acknowledgments}

\nocite{*}
\bibliography{Windrush_bib}



\end{document}